%% file: model.tex
\pdfoutput=1

\documentclass[longauth]{aa}

\usepackage{comment}
\usepackage{graphicx}
\usepackage{multirow}
\usepackage{booktabs}
\usepackage{textgreek}
\usepackage{xargs}
\usepackage[dvipsnames]{xcolor}
\usepackage{xspace}
\usepackage{caption}
\usepackage{cuted}
\usepackage{capt-of}
\usepackage{subcaption}
\usepackage{float}
\usepackage{enumitem}
\usepackage[rightcaption]{sidecap}
\usepackage{txfonts}
\usepackage{hyperref}
\usepackage{amsmath}
\hypersetup{
    colorlinks=true,
    linkcolor=blue,
    citecolor=blue,
    filecolor=magenta,
    urlcolor=cyan,
}
\usepackage{subcaption}
\usepackage{adjustbox}

\graphicspath{{./}{figs/}}
\input{config/macros}

\input{config/emlines}
\defcitealias{Cataldi2025}{C25}
\defcitealias{Cataldi2026}{C26}

\begin{document}

\titlerunning{Abundance Patterns at high redshift}
\authorrunning{E. Cataldi et al.}

\title{Chemical abundance patterns at $z = 1-7$: Carbon, Nitrogen, Neon, Sulphur, and Argon abundances from JWST spectroscopy}

\author{
E. Cataldi\inst{\ref{UNIFI}, \ref{arcetri}} \and
F. Belfiore\inst{\ref{arcetri}, \ref{eso}} \and
M. Curti\inst{\ref{oas}} \and
B. Moreschini\inst{\ref{UNIFI},\ref{arcetri}} \and
A. Marconi\inst{\ref{UNIFI},\ref{arcetri}} \and
F. Mannucci\inst{\ref{arcetri}} \and
G. Cresci\inst{\ref{arcetri}} \and
A. Feltre\inst{\ref{arcetri}} \and
M. Ginolfi\inst{\ref{UNIFI},\ref{arcetri}} \and
A. Amiri\inst{\ref{IPM},\ref{arkansas}} \and
M. Arnaboldi\inst{\ref{eso}} \and
S. Bhattacharya\inst{\ref{herts}}\and
E. Bertola\inst{\ref{arcetri}}\and
C. Bracci\inst{\ref{UNIFI},\ref{arcetri}}\and
M. Ceci\inst{\ref{UNIFI},\ref{arcetri}}\and
A. Chakraborty\inst{\ref{IIA}}\and
Q. D'Amato\inst{\ref{iaps}} \and
C. Kobayashi\inst{\ref{herts}}\and
C. Marconcini\inst{\ref{arcetri}}\and
R. Maiolino\inst{\ref{cavensish}, \ref{kicc}} \and
L. Querci\inst{\ref{UNIFI}} \and
M. Scialpi\inst{\ref{trento},\ref{arcetri},\ref{UNIFI}}\and
L. Ulivi\inst{\ref{trento},\ref{arcetri},\ref{UNIFI}} \and
G. Venturi\inst{\ref{arcetri}}
\and
M. V. Zanchettin\inst{\ref{arcetri}, \ref{iaps}} 
}

\institute{
\label{UNIFI} Università di Firenze, Dipartimento di Fisica e Astronomia, via G. Sansone 1,
  50019 Sesto Fiorentino, Florence, Italy \and
\label{arcetri} INAF -- Osservatorio Astrofisico di Arcetri, Largo E. Fermi 5, I-50125,
  Florence, Italy \and
\label{oas} INAF -- Osservatorio di Astrofisica e Scienza dello Spazio di Bologna, Via Gobetti, 93/3, I-40129 Bologna, Italy \and
\label{eso} European Southern Observatory, Karl-Schwarzschild Straße 2, D-85748 Garching
  bei München, Germany \and
\label{IPM} School of Astronomy, Institute for research in fundamental sciences (IPM),
  Tehran, P.O. Box 19395-5531, Iran \and
\label{arkansas} Department of Physics, University of Arkansas, 226 Physics Building,
  825 West Dickson Street, Fayetteville, AR 72701, USA \and
\label{herts} Centre for Astrophysics Research, Department of Physics, Astronomy and
  Mathematics, University of Hertfordshire, Hatfield AL10 9AB, UK \and
\label{cavensish} Cavendish Laboratory, University of Cambridge, 19 JJ Thomson Avenue,
  Cambridge, CB3 0HE, UK \and
\label{kicc} Kavli Institute for Cosmology, University of Cambridge, Madingley Road,
  Cambridge, CB3 0HA, UK \and
\label{trento} University of Trento, Via Sommarive 14, I-38123 Trento, Italy \and
\label{iaps} INAF -- Istituto di Astrofisica e Planetologia Spaziali, Via Fosso del Cavaliere 100, 00133 Rome, Italy
\and 
\label{IIA} Indian Institute of Astrophysics, 2nd Block, Koramangala, Bangalore 560 034, India
}

\date{}

\abstract{
The relative abundances of chemical elements in the interstellar medium (ISM) trace the history of star formation and chemical enrichment in galaxies. We investigate the abundance patterns of carbon (C), nitrogen (N), neon (Ne), sulfur (S), and argon (Ar) in star-forming galaxies at $z\sim1$–7, combining ultra-deep JWST/NIRSpec observations from the MARTA survey with archival spectroscopy. We assemble 244 galaxies with \OIIIopt$\lambda4363$ detections and derive direct $T_e$-based abundances. New empirical strong-line calibrations for N/O, Ne/O, S/O, and Ar/O extend the analysis to an additional 514 galaxies.
We find distinct abundance–metallicity trends. N/O is enhanced by $\sim0.18$ dex relative to local galaxies, with a stronger enhancement at low metallicity. Direct measurements reveal decreasing S/O and Ar/O with increasing metallicity, while strong-line measurements suggest a flattening at higher metallicities, especially for S/O. We also identify a stellar-mass dependence in the O/Ar–Ar/H plane, with lower-mass galaxies exhibiting lower O/Ar at fixed Ar/H, consistent with differences in star-formation efficiency and delayed SNe Ia enrichment. Ne/O remains approximately constant up to $12+\log(\mathrm{O/H})\sim8.2$, with a mild increase at higher metallicities and no significant redshift evolution. C/O increases with metallicity, broadly following the local relation.

Finally, a forward model shows that observational selection affects the abundance ratios differently: it mainly restricts the accessible metallicity range for Ne/O and Ar/O, but preferentially selects galaxies with enhanced N/O at low metallicity, with a weaker effect for S/O. These results provide new constraints on chemical enrichment at high redshift and highlight the importance of accounting for selection effects when interpreting abundance–metallicity trends.
}


\maketitle


\section{Introduction}
\label{sec:intro}

The build-up of heavy elements provides one of the most direct observational windows into the processes that regulate galaxy evolution. Metals trace the cumulative effects of star formation (SF), gas accretion, and stellar feedback, while simultaneously shaping the physical properties of the interstellar medium (ISM) by regulating cooling, SF efficiency, dust content and opacity \citep[e.g.,][]{sutherland_cooling_1993, krumholtz_2012, bolatto_co--h2_2013, Draine_2011, maiolino_re_2019, peroux_2020}. Following the evolution of chemical abundances across cosmic time is therefore key to reconstructing the baryon cycle in galaxies.

Given that oxygen is the most abundant metal in mass, its overall abundance is often used as a tracer of the integrated enrichment history. Relative abundance ratios of different elements, instead, offer a differential probe of the underlying nucleosynthetic channels. 
Different elements are, in fact, produced by stars of different masses and on different timescales, making their ratios sensitive to the star formation history (SFH), the initial mass function (IMF), and the occurrence of inflows and outflows \citep[e.g.,][]{maiolino_re_2019, curti_2026_review}. 
In massive stars ($>8~M_{\odot}$) oxygen is efficiently released into the ISM through stellar winds and core-collapse supernovae (CCSNe) on short timescales of $\approx$ 3--40~Myr \citep[e.g.,][]{Woosley2002, Chiappini2003, Limongi_Chieffi_2018, kobayashi_origin_2020}. Under typical \hii region conditions, the total oxygen abundance is well approximated by the sum of O$^+$ and O$^{2+}$, with higher ionisation stages usually providing negligible corrections \citep[e.g.,][]{stasinska_abundance_2002}. The emission lines associated with these ionic species lie in the rest-frame optical, providing a rich set of diagnostics for oxygen abundances; O/H is therefore commonly adopted as a proxy for the gas-phase metallicity of the ISM \citep{kewley_understanding_2019}.

The nitrogen-to-oxygen ratio (N/O) is one of the most extensively studied abundance ratios in the ISM of star-forming galaxies, as the emission lines of nitrogen are also readily accessible in the same optical spectral range as the prominent oxygen lines. Nitrogen can be produced through two main channels, referred to as primary and secondary production. In the primary channel, nitrogen originates from carbon and oxygen freshly synthesized within the star itself: these newly formed nuclei are mixed into the hydrogen-burning regions and converted into nitrogen through the CNO cycle. As a result, the nitrogen yield in this case is largely independent of the star’s initial metallicity. Primary nitrogen can be produced both by massive stars on short timescales--tens of Myr--where fast rotational and convective mixing processes are highly efficient \citep[e.g.,][]{2000ApJ...541..660H,  2002A&A...390..561M,chiappini_stellar_rotation_2006}, and by low- and intermediate-mass stars (LIMS hereafter; $M\lesssim8 M_\odot$), during their asymptotic giant branch (AGB) phase \citep[e.g.,][]{vincenzo_nitrogen_2016}. The latter happens on much longer enrichment timescales--hundreds of Myr to several Gyr. 

In contrast, the secondary channel for N production relies on C and O already present in the progenitor gas cloud. In this case, the CNO cycle transforms pre-existing heavy elements into nitrogen, leading to a metallicity-dependent yield \citep[e.g.,][]{2000ApJ...541..660H, vincenzo_nitrogen_2016, maiolino_re_2019}. This contribution arises mainly from LIMS in their AGB phase \citep{kobayashi_isotopes_MW_2011, nomoto_chemical_evolution_2013, ventura_agb_yields_2013}.
Both channels thus operate over a wide range of stellar masses and timescales \citep[e.g.,][]{vincenzo_nitrogen_2016, maiolino_re_2019}. The N/O ratio is therefore a sensitive clock of chemical enrichment and has attracted considerable attention in recent JWST studies of high-$z$ galaxies \citep[e.g.,][]{bunker_gnz11_2023, Ji_GNz11_2024, Rogers2024_CECILIA, Arellano-Cordova25b,  Curti_2025_gsz9, 2025ApJ...982...14H, Napolitano_2025, Navarro_Carrera_2025, 2025ApJ...981..136S, 2025ApJ...980..225T, Berg2026, Cataldi2026, Cameron2026, Morel_2026_nitrogen, Ji2026, Schaerer2026, Zhang2026}.

Carbon plays a crucial role in the chemical evolution of galaxies, along with oxygen and nitrogen. It is produced in the interior of stars through the triple-$\alpha$ process
during helium burning, a process that is independent of the star's initial metallicity. Carbon can be released into the ISM through two main pathways, with by different delay times. LIMS produce carbon during He burning through the triple-$\alpha$ process and during thermal pulses
in the AGB phase. The newly produced C is transported to the stellar envelope through third dredge-up episodes and subsequently released into the ISM through stellar winds. The carbon yield generally peaks for stellar masses around $2$--$4\,M_\odot$ and decreases towards higher masses because hot-bottom burning
(HBB) converts C into N \citep{kobayashi_isotopes_MW_2011, Karakas2014}. Massive stars (M > 8 M$_{\odot}$), on the other hand, release carbon on shorter timescales (a few Myr), primarily through CCSNe explosion or possibly through the strong stellar winds of the Wolf-Rayet (WR) phase \citep[e.g.,][]{Woosley2002, 2002A&A...390..561M}.

Finally, Ne, S, and Ar are all $\alpha$-elements: their most abundant isotopes have mass numbers that are multiples of four, and they are synthesised through $\alpha$-capture reactions in massive stars on the same timescales (up to few tens Myr) as oxygen \citep[e.g.,][]{Izotov2006, croxall_chaos_2016, esteban2020carbon, rogers2022chaos, 2025MNRAS.540.2991A}.
If this was the only nucleosynthetic channel, their abundance ratios would remain approximately constant with metallicity. 
However, nucleosynthesis models also predict that Type~Ia supernovae (SNe~Ia) contribute non-negligibly to the S and Ar budget \citep[e.g.,][]{Iwamoto1999, Johnson2019, kobayashi_origin_2020}. This contribution may manifest itself as an observable trend of S/O and Ar/O with metallicity.
The nucleosynthetic time delay of $\sim 100 \, \rm{Myr}-Gyr$ between CCSNe and SNe~Ia opens up the possibility of using O/Ar (and, possibly, O/S) as analogues of the classical [$\alpha/\text{Fe}$] diagnostic -- a well-established tracer of SF timescales in resolved stellar populations \citep[e.g.,][]{Tinsley1979, Matteucci1990, Tolstoy2009, Matteucci2021}. In this framework, the ISM abundance ratio between an $\alpha$-element produced predominantly in CCSNe (such as oxygen) and one with a non-negligible SNe~Ia contribution (such as argon or sulfur) carries an imprint of the star formation history of the host system. 

In the local Universe, most observational studies find that Ne/O, S/O, and Ar/O are broadly constant with metallicity, within the scatter and measurement uncertainties \citep[e.g.,][]{kennicutt_composition_2003, izotov_chemical_2006, croxall_chaos_2016, arellano_cordova_2020_abundances,berg_chaos_2020, rogers2022chaos, arellano_cordova_2024_CLASSY, Esteban2025_alpha_elements}, while C/O rises with oxygen abundance \citep[e.g.,][]{berg_carbon_2016, Berg_CNO_2019}.
However, several works report mild trends with metallicity for S/O and Ar/O for specific samples \citep[e.g.,][]{Dors2016, Diaz_Zamora_2022, Arnaboldi2022, Brazzini2024, Scholte2026_elements}, suggesting that the constancy of these abundance ratios with metallicity is not firmly established. 

At high-$z$, the situation is even less constrained: the nebular emission lines used for tracing some of these elements (e.g. \ArIII$\lambda$7136, \NeIII$\lambda$3869, \SII$\lambda\lambda$6716,6730, and \SIII$\lambda\lambda$9069,9532),
have been detected in distant galaxies only thanks to the sensitivity of the Near InfraRed Spectrograph (NIRSpec) on board JWST. Early results, based on samples of only a dozen galaxies, suggest that the Ne/O abundance ratio remains broadly consistent with the local Universe, showing little or no evidence for redshift evolution \citep{arellano-corodva_2023, 2025MNRAS.540.2991A, isobe_CNO_2023, Stanton_excels_2025}.  
The situation is less clear for Ar and S. \citet{Bhattacharya2025_OAr} first used the O/Ar--Ar/H plane to investigate the relative contributions of CCSNe and SNe~Ia to the chemical enrichment of high-$z$ star-forming galaxies. Later studies reported predominantly subsolar Ar/O ratios, consistent with a limited contribution from SNe~Ia relative to CCSNe \citep{Stanton_excels_2025, Rogers2026_CECILIA, Foley2026_Argon_AURORA}. \citet{Rogers2026_CECILIA} also found subsolar S/O, which they interpreted within the same framework of predominantly CCSN-driven enrichment.
Overall, existing studies are based on small samples, highlighting the need for larger statistics to robustly establish the evolution of abundance ratios across cosmic time.

In this work, we exploit the capabilities of JWST to assemble a spectroscopic sample of 244 star-forming galaxies at $z\sim1$--7 with direct metallicity measurements based on the temperature-sensitive \OIIIopt$\lambda$4363 line. Within this parent sample, we identify subsamples with detections of emission lines tracing C, Ne, N, S, or Ar. Building on the observational framework established in \citet{Cataldi2026} (\citetalias{Cataldi2026} hereafter) for nitrogen, we extend the analysis to the determination of direct-method abundances for Ar, C, S and Ne. We further present calibrations for N/O, S/O, Ar/O and Ne/O based on strong line ratios. With this large sample of abundance measurements at high redshift, we aim to shed new light on the behaviour of abundance trends as a function of metallicity.

The paper is structured as follows. Section~\ref{sec:data} describes the observational dataset. Section~\ref{sec:measuring} presents the abundance measurement methods. Section~\ref{sec:calibrations} derives and discusses the strong-line calibrations, including an investigation of their secondary dependencies. Section~\ref{sec:abundances} presents the observed abundance patterns in the different X/O--O/H planes and compares them with galactic chemical evolution (GCE) models and local relations. Section~\ref{sec:discussion} explores possible physical explanations for the observed trends and assesses the impact of observational selection effects. Finally, Section~\ref{sec:conclusions} summarises our main findings.

\section{Observational data}
\label{sec:data}

\begin{figure}
    \centering
    \includegraphics[width=\linewidth]{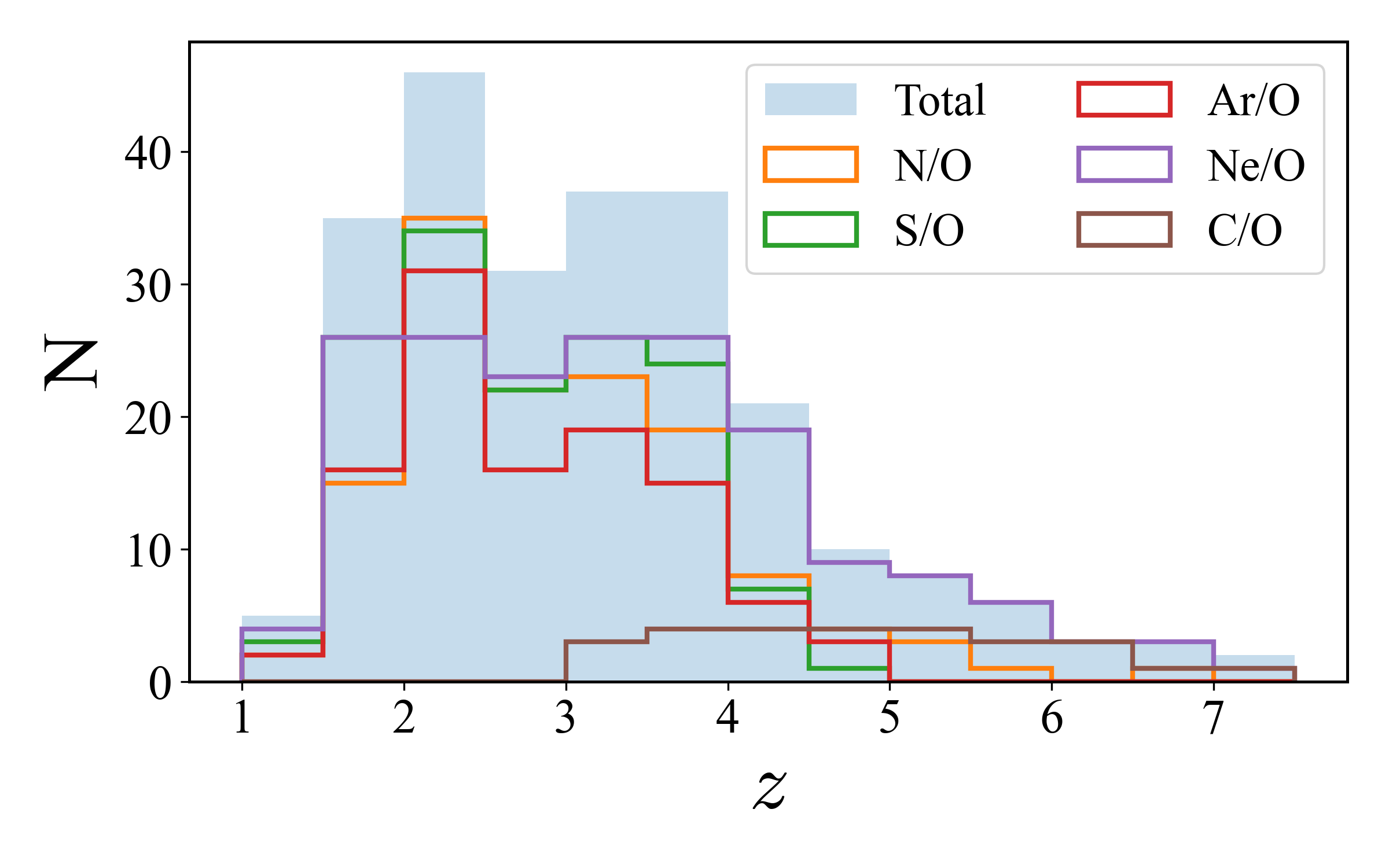}
    \caption{Redshift distribution of the total sample of high-z redshift galaxies for which we were able to directly measure the oxygen abundance (namely, 244 object) compared with the subsamples for which it was also possible to infer the respective abundance ratio: N/O (in orange), S/O (in green), Ar/O (in red), Ne/O (purple) and C/O (in brown).}
    \label{fig:z_distribution}
\end{figure}

We assembled a sample of star-forming galaxies at $z \sim 1-7$ with JWST/NIRSpec spectroscopy. This work builds upon the work of \citetalias{Cataldi2026}, which focused on nitrogen, and substantially expands to include Ar, C, S and Ne. 
We selected objects with a signal-to-noise ratio $\mathrm{S/N} > 5$ in both \Halpha\ and \Hbeta\ to ensure a robust $\Halpha/\Hbeta$ ratio and avoid introducing large uncertainties in the dust attenuation correction.
Depending on the diagnostic of interest, we subsequently imposed a threshold of $\mathrm{S/N} > 3$ on the other emission lines of interest: \CIII$\lambda\lambda$1907,1909, \OII$\lambda\lambda$3727,3729, \NeIII$\lambda$3869, \OIIIopt$\lambda$5007, \NII$\lambda$6583, \SII$\lambda\lambda$6716,6730, and \ArIII$\lambda$7135. When using the T$_{\rm{e}}$ method, we additionally imposed $\mathrm{S/N} > 3$ on the auroral \OIIIopt$\lambda$4363 line. Not all objects in the sample have simultaneous detections of all emission lines of interest. Therefore, the number of galaxies contributing to each abundance ratio measurement varies accordingly. As a result, the subsamples used for determining direct abundances of N/O, Ar/O, C/O, S/O and Ne/O analyses are not identical, and their sizes are reported in Table~\ref{tab:bestfits}. The redshift distribution of each subsample is instead illustrated in Figure \ref{fig:z_distribution}.

AGN contamination was assessed and excluded using the standard emission line diagnostics, namely \OIIIopt/\Hbeta versus \NII/\Halpha adopting the redshift-dependent calibration of \citet{kewley_cosmic_2013}. We also tested the diagnostic of \citet{2024A&A...691A.345M} based on \OIIIopt$\lambda$4363. This diagnostic yields a substantially different classification for our sample: approximately 35\% of the galaxies classified as star-forming according to \citet{kewley_cosmic_2013} fall within the AGN region of the \OIIIopt$\lambda$4363-based diagnostic (although most lie in the proximity of the classification boundary). Nevertheless, we verified that this difference in sample selection does not affect the observational results of this work: although the alternative selection results in a smaller sample, all the observational trends of chemical abundance patterns remain clearly present. We therefore adopt the widely established \citet{kewley_cosmic_2013} diagnostic as our fiducial classification, while noting that our main conclusions are robust to the choice of AGN diagnostic.

\subsection{The MARTA survey}
\label{sec:MARTA}

The Measuring Abundances at high Redshift with the $T_{\rm e}$ Approach (MARTA; PID 1879, PI Curti) programme is a deep JWST/NIRSpec Micro Shutter Array (MSA) campaign targeting star-forming galaxies at Cosmic Noon ($z \sim 2$--$3$). A key goal of the programme is the detection of the temperature-sensitive auroral line \OIIIopt$\lambda$4363 with the G140M grating ($R \sim 1000$), enabling robust direct-method metallicity determinations. The total integration time is 31.9 hours with G140M and 7.4 hours with G235M. The MARTA survey is described in detail in \citet[][hereafter \citetalias{Cataldi2025}]{Cataldi2025}. 
We obtained approximately 110 spectra of galaxies with confirmed spectroscopic redshift.

\subsection{JWST direct sample}
\label{sec:DAWN}

We supplemented the MARTA data with a compilation of galaxies at $z > 1$ drawn from 
the literature, selecting additional galaxies which satisfy the same criteria in terms of emission line selection and S/N thresholds, obtained from the publicly available JADES DR4 catalogue \citep{CurtisLake_2026_JADES_DR4, Scholtz_2026_JADES}, and from the DAWN JWST Archive v4.4 \citep[DJA;][]{2025A&A...699A.358V}\footnote{Repository website: \url{https://dawn-cph.github.io/dja/}}, a public repository of reduced JWST NIRSpec MSA spectroscopy processed with MSAExp \citep{Brammer2023}, from which we adopted the standard emission-line flux catalogues directly\footnote{JADES galaxies are excluded from the DJA catalogue of emission lines to avoid duplication.}. 
For all galaxies drawn from JADES DR4 and the DJA, the emission-line fluxes used here were measured from JWST/NIRSpec medium-resolution grating spectra (G140M, G235M, and G395M).
The main programs used in this work are summarised in Appendix \ref{app:jwst_programs}.

We also compiled published oxygen, sulfur, argon and neon abundances of galaxies at cosmic noon from recent works of
\citet{Rogers2026_CECILIA, Stanton_excels_2025} and \citet{Foley2026_Argon_AURORA}, which we use for comparison with our analysis.

\subsection{JWST strong-line sample}
\label{sec:strong_line_sample}

In addition to the direct-method sample described above, we assembled a larger sample of 514 galaxies satisfying the same selection criteria, except for the auroral-line detection requirement. This sample includes galaxies from MARTA, JADES, and the DJA with S/N above the adopted thresholds for all relevant emission lines. For these galaxies, we derived chemical abundances using the strong-line calibrations described in Sect.~\ref{sec:strong_line_method}.

\subsection{Local calibration sample}
\label{sec:local}

A local calibration sample is used to give a comparative view of all the main results presented in Sections \ref{sec:calibrations} and \ref{sec:abundances}. It comprises low-redshift galaxies \citep{vale_asari_bond_2016, curti_new_2017,Khoram_2025} and \hii\ regions \citep{2009A&A...503...61I, 2009A&A...505...63G, guseva_vlt_2011,berg_chaos_2015, berg_chaos_2020, rogers_chaos_2021,2021MNRAS.500.2359Z, rogers2022chaos} with direct-method oxygen, nitrogen, sulphur, argon abundances. A detailed description of the various subsamples is presented in \citetalias{Cataldi2025} and \citetalias{Cataldi2026}. 
For all of these objects, ionic abundances were derived self-consistently by recovering the published line fluxes and recomputing electron temperatures following the methodology of Section~\ref{subsec:direct_method}. This local sample spans a wide range in metallicity (12+log(O/H) $\in [7,9]$) and constitutes the empirical anchor for the strong-line calibrations.

For carbon, we instead compiled published abundance measurements directly from the literature. Specifically, we included a sample of 10 nearby metal-poor star-forming galaxies observed with HST \citep{senchyna_2017}, 11 additional LyC-leaking galaxies observed with HST \citep{Izotov_2023_LyC_galaxies}, and 32 nearby low-metallicity, high-ionization dwarf galaxies observed with HST \citep{berg_carbon_2016, Berg_CNO_2019}, as well as carbon and oxygen abundances derived from recombination-line measurements in giant extragalactic \hii regions \citep{Esteban_2002_RL_abundances, esteban_keck_2009}. Finally, we included measurements for Galactic and extragalactic \hii regions from \citet{Garcia_Rojas_2007_ADF}, and recombination-line abundances for \hii regions in the blue compact dwarf galaxy NGC 5253 obtained with VLT/UVES \citep{Lopez_Sanchez_2007}.

\section{Chemical abundance measurements}
\label{sec:measuring}

\subsection{Dust correction}
\label{subsec:dust_corr}

Observed emission-line fluxes of all the samples described in Section \ref{sec:data} are corrected for dust attenuation using the Balmer decrement. 
For each object, we fit all Balmer lines detected at $\mathrm{S/N}>5$ (i.e. a subset of \Halpha, \Hbeta, \Hgamma, and \Hdelta depending on the specific target) to their theoretical Case~B recombination values, assuming $T_{\rm e}=10^4$ K and $n_{\rm e}=100$ cm$^{-3}$, representative of the physical conditions in typical \hii regions and commonly adopted \citep{osterbrock_astrophysics_2006}. We adopt the \citet{cardelli_relationship_1989} extinction law with $R_V=3.1$ to derive the color excess $E(B-V)$ via a global $\chi^2$ minimisation.

As a consistency check, we also explored a simultaneous Markov Chain Monte Carlo (MCMC) fit of the dust attenuation and gas physical properties. In this approach, $E(B-V)$ is constrained by the observed Balmer-line ratios, while $T_{\rm e}$ and $n_{\rm e}$ are constrained independently by the available diagnostic lines: auroral lines for $T_{\rm e}$ and the \SII$\lambda6716/\lambda6731$ doublet for $n_{\rm e}$. When these diagnostics are not available, $T_{\rm e}$ and $n_{\rm e}$ are fixed to their fiducial values. We found that the resulting $E(B-V)$ values change negligibly with varying physical conditions, with variations of less than 2\%. Since this approach can only be applied to the galaxies with all the required diagnostic lines, we do not adopt it as our fiducial attenuation-correction method.

Employing alternative extinction laws, such as \citet{calzetti_dust_1994} or the Large Magellanic Cloud law \citep{gordon_LMC_attenuation_2003}, produces negligible differences, on average of $\sim$12\% on the computed $E(B-V)$ which translates in changes of less than $1\%$ both on inferred temperatures and metallicities.

We also tested the impact of adopting a high-redshift nebular attenuation curve, as recently proposed by \citet{Reddy_2026_attenuation}, using their median attenuation law (their Eq. 22). This prescription and our fiducial \citet{cardelli_relationship_1989} law yield very similar colour excesses, with a median relative offset of $\sim-10\%$. This difference has a negligible impact on the derived physical conditions and abundances, again resulting in variations of less than 1\% in temperatures and abundances. It is also worth noting that $\approx45\%$ of our targets have $E(B-V)=0$, limiting the impact of the adopted attenuation curve on our results.

\subsection{Direct method}
\label{subsec:direct_method}

For our sample of  galaxies with auroral line detections (Sec. \ref{sec:MARTA}, \ref{sec:DAWN}, \ref{sec:local}), ionic abundances were derived using the $T_{\rm e}$ method, which exploits the temperature sensitivity of collisionally excited emission lines: electron temperatures are inferred from auroral-to-nebular line ratios and then used to determine the ionic abundances.

The final products are $12 + \log(\mathrm{O/H})$ from oxygen lines, and the abundance ratios $\log(\mathrm{N/O})$, $\log(\mathrm{Ar/O})$, $\log(\mathrm{C/O})$, $\log(\mathrm{S/O})$, and $\log(\mathrm{Ne/O})$ from the respective line ratios:
\begin{itemize}
\item{N2O2 = $\log(\NII\,\lambda6584/\OII\,\lambda\lambda3727, 29)$, }
\item{Ar3O3 =$\log(\ArIII\,\lambda7135/\OIIIopt\,\lambda5007)$, }
\item{C3O3 =$\log(\CIII\,\lambda\lambda1907,1909/\OIIIopt\,\lambda5007)$, }
\vspace{1mm}
\item{
S23/R23 $= \log \frac{
\SII \,\lambda\lambda 6717,6731 + \SIII \,\lambda\lambda 9071,9533
}{
\OII \,\lambda\lambda 3727,3729 + \OIIIopt \,\lambda 5007
}$
}\vspace{1mm}
\item{Ne3O3 = $\log(\NeIII\,\lambda3869/\OIIIopt\,\lambda5007)$.}
\end{itemize}
each corrected via the appropriate ICF.

\subsubsection{Electron density and temperatures}
\label{subsec:Te}

The electron density $n_{\rm e}$ is derived from the \SII$\lambda\lambda$6716,6730 doublet ratio when both lines are available at S/N>3, which is the case for $\sim$50\% of our high-$z$ targets. For the remaining objects, a fiducial value of $n_{\rm e} = 400\,\mathrm{cm}^{-3}$ is adopted, corresponding to the median density measured for targets in our sample with a detected \SII doublet and consistent with the typical values reported for high-$z$ galaxies \citep[e.g.,][]{sanders_mosdef_2016, Stanton_excels_2025, Topping_2025_density}.
As a consistency check, for targets in which the \SII doublet is not detected at $3\sigma$, we also computed redshift-dependent densities using the parametrisation from \citet{Topping_2025_density}. Adopting this redshift-dependent parametrisation instead of the fiducial value changes the resulting electron temperatures by less than 1\% and the inferred metallicities by less than 0.1\%, so we retain the fiducial value throughout.

Ionic abundances are computed from dust-extinction-corrected fluxes, $T_{\rm e}$, and $n_{\rm e}$ using the \texttt{PyNeb} \textsc{getIonAbundance} function \citep{luridiana_pyneb_2015}. Specifically, the dust-corrected line fluxes are converted into ionic abundances by dividing by the corresponding emissivities, which depend on the adopted $T_{\rm e}$, and $n_{\rm e}$. The electron temperature assigned to each ionic species is guided by the structure of the ionisation zones, although the mapping is not always one-to-one. A summary of the lines used and the atomic data adopted is provided in Appendix \ref{app:atomic}.

In particular, we derive the electron temperature of the high-ionisation zone, $T_{\rm e}(\mathrm{O}^{2+})$, from the \OIIIopt$\lambda$4363/\OIIIopt$\lambda$5007 auroral-to-nebular line ratio. When both \SII lines are detected at S/N >3, we simultaneously constrain $T_{\rm e}(\mathrm{O}^{2+})$ and $n_{\rm e}$  using the \texttt{getCrossTemDen} routine in \texttt{PyNeb}, which takes the \OIIIopt$\lambda$4363/\OIIIopt$\lambda$5007 and \SII$\lambda$6731/\SII$\lambda$6716 ratios as input and return both the electron temperature and density; otherwise, we derive $T_{\rm e}(\mathrm{O}^{2+})$ from the \OIIIopt aur/neb ratio alone using \texttt{getTemDen}, adopting the density constraint described above. 

For the low-ionisation zone, we assume that N$^+$, S$^+$, and O$^+$ arise mostly from a common region, again motivated by the proximity of their ionisation potentials. We accordingly adopt a single electron temperature, $T_{\rm e}(\mathrm{O}^{+})$, to derive $\mathrm{N^{+}/H^+}$  and $\mathrm{S^{+}/H^+}$ as well. When the \OII$\lambda\lambda$7319,7329 auroral lines are detected at S/N>3 ($\sim$5\% of our targets), $T_{\rm e}(\mathrm{O}^{+})$ is derived directly from their auroral-to-nebular ratio; for the remaining targets, we instead infer $T_{\rm e}(\mathrm{O}^{+})$ from the fiducial $T_{\rm e}(\mathrm{O}^{+})$--$T_{\rm e}(\mathrm{O}^{2+})$ relation of \citetalias{Cataldi2025}, calibrated on high-redshift systems.


The case of intermediate-ionisation species such as $\mathrm{S}^{2+}$, $\mathrm{Ar}^{2+}$ and $\mathrm{C}^{2+}$ is less straightforward, as their ionisation zones partially overlap with both $\mathrm{O}^+$ and $\mathrm{O}^{2+}$. We therefore tested two different approaches to estimate their ionic abundances: one adopting the directly measured $T_{\rm e}(\mathrm{O}^{2+})$, and the other using $T_{\rm e}(\mathrm{S}^{2+})$ as a proxy for the electron temperature of the intermediate-ionisation zone. Since $T_{\rm e}(\mathrm{S}^{2+})$ is not directly constrained by auroral-line measurements in our sample, we infer it from empirical calibrations from $T(\mathrm{O}^{2+})$ using the relation presented in \citet{garnett_electron_1992}, which has recently been shown by \citet{Rogers2026_CECILIA} to provide an accurate description of cosmic-noon galaxies based on direct temperature measurements from the CECILIA \citep{Strom2023CECILIA:} and AURORA samples.
Applying this calibration, we find that S, Ar and C abundances derived using $T_{\rm e}(\mathrm{S}^{2+})$ are systematically higher than those obtained with $T_{\rm e}(\mathrm{O}^{2+})$, with a median offset of 0.09 dex. Nevertheless, although the absolute abundance scale is shifted, the observed trends remain unchanged and all the main conclusions of this work are unaffected. For this reason, we adopt $T_{\rm e}(\mathrm{O}^{2+})$ throughout the paper. While this choice may introduce a systematic offset in the absolute $\mathrm{Ar}^{2+}$ , $\mathrm{S}^{2+}$ , and $\mathrm{C}^{2+}$ abundances, it relies on directly measured temperatures rather than on empirically inferred ones, thereby avoiding the additional uncertainties associated with calibration-based estimates.

For neon, $\mathrm{Ne}^{2+}$ and $\mathrm{O}^{2+}$ predominantly coexist in the highly ionised inner regions of the nebula, although their ionisation zones are not strictly coincident (see Table in Appendix~\ref{app:atomic}). $\mathrm{Ne}^{2+}$ preferentially traces somewhat more highly ionised gas than $\mathrm{O}^{2+}$, placing it between $\mathrm{O}^{2+}$ and higher-ionisation species such as $\mathrm{C}^{3+}$ and $\mathrm{N}^{3+}$. The choice of the appropriate electron temperature is therefore not straightforward, since neither direct auroral-line constraints nor widely established empirical calibrations of $T_{\rm e}$ for $\mathrm{C}^{3+}$ and $\mathrm{N}^{3+}$ are currently available. Given the substantial overlap between the $\mathrm{Ne}^{2+}$ and $\mathrm{O}^{2+}$ ionisation zones, we therefore adopt $T_{\rm e}(\mathrm{O}^{2+})$ for $\mathrm{Ne}^{2+}$.

\subsubsection{Ionisation correction factors}
\label{subsec:ICFs}

Ionisation correction factors (ICFs) are applied to convert ionic abundances into total elemental abundances by accounting for unobserved ionisation stages. We adopt the prescriptions of \citet{amayo_ICFs_2020}, derived from a large grid of photoionisation models selected to reproduce the properties of 985 giant \hii\ regions in spiral galaxies and blue compact galaxies. To assess the sensitivity of our results to this choice, we recalculated the abundances using the widely adopted ICFs of \citet{izotov_chemical_2006}. The two prescriptions give broadly consistent results: the median offsets are $\lesssim 0.06$ dex for all abundance ratios except Ne/O, for which the offset is approximately 0.1 dex. A similar difference for Ne/O was reported by \citet{Esteban2025_alpha_elements}, who found mean values differing by 0.12 dex between the Amayo and Izotov prescriptions in a sample of local star-forming regions. 
In all cases, however, the overall abundance trends remain unchanged.

\subsection{Strong-line method}
\label{sec:strong_line_method}

For the high-$z$ sample with no auroral line detections (Sec.~\ref{sec:strong_line_sample}) we inferred chemical abundances using empirical strong-line calibrations. We determined oxygen abundances by inverting the high-redshift metallicity calibrations of \citetalias{Cataldi2025} through $\chi^2$ minimisation, following the methodology described in \citetalias{Cataldi2026}. We then inferred the other abundance ratios using the strong-line relations calibrated in Section~\ref{sec:calibrations}.

\section{Strong-line calibrations for N, Ar, S, and Ne}
\label{sec:calibrations}


\begin{figure*}
    \centering
    \sidecaption
    \includegraphics[width=12cm]{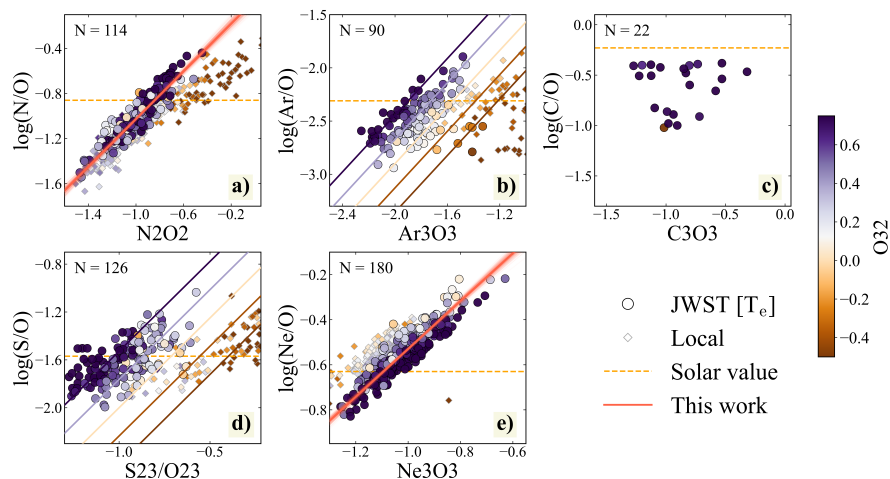}
    \caption{Calibrations of the abundance ratios ($\mathrm{N/O}$, $\mathrm{Ar/O}$, $\mathrm{C/O}$, $\mathrm{S/O}$, and $\mathrm{Ne/O}$) as a function of N2O2, Ar3O3, C3O3, S23/R23 and Ne3O3, respectively. Circles represent the combined high-redshift sample, namely MARTA (Sect. \ref{sec:MARTA}) DJA and JADES DR4 samples (Sect.~\ref{sec:DAWN}), while small diamonds show the local reference sample (Sect.~\ref{sec:local}). For the Ar/O panel we also include objects from \cite{Foley2026_Argon_AURORA}.
    For $\mathrm{N/O}$ and $\mathrm{Ne/O}$, the solid red line and the surrounding shaded bundle illustrate the best-fit linear relation and its $1\sigma$ Monte Carlo uncertainties. For $\mathrm{Ar/O}$ and $\mathrm{S/O}$ the coloured lines represent the fixed 3D empirical relations evaluated at specific $\mathrm{O32}$ values via a linear two-parameter fit (see colour bar). No fit is presented for C/O.}
    \label{fig:strong_lines}   
\end{figure*}

\begin{table}
\caption{Best-fit parameters of the abundance calibrations.}
\label{tab:bestfits}
\centering
\begin{tabular}{lccccccc}
\hline\hline
 & Tracer & $N$ & $\langle z \rangle$ & $\alpha$ & $\beta$ & $\gamma$ & $\sigma_{\rm int}$ \\
\hline
N/O & N2O2 & 114 & 2.8 & $0.03$  & $1.06$ & --            & 0.07 \\
 &  & & & $\pm0.06$ & $\pm0.04$ &  & \\
Ar/O  & Ar3O3  & 90  & 2.6 & $-0.24$ & $1.33$ & $0.57$ & 0.05 \\
  &  & & & $\pm0.07$ & $\pm0.04$ & $\pm0.03$ &  \\
S/O  & $\frac{\rm S23}{\rm R23}$ & 126 & 2.6 & $-0.46$ & $1.53$ & $0.59$  & 0.08 \\
   & &     &        & $\pm0.09$ & $\pm0.12$ & $\pm 0.07$ &   \\
Ne/O  & Ne3O3   & 180 & 3.3 & $0.53$  & $1.06$ & -- & 0.04 \\
    &  & & & $\pm0.04$  & $\pm0.04$ & -- &  \\
\hline
\end{tabular}

\medskip
\parbox{0.96\linewidth}{\footnotesize
Columns report the number of galaxies with direct abundance measurements used in each fit ($N$), the best-fit coefficients of the relation
$\log(X/{\rm O}) = \alpha + \beta\,Y + \gamma\,{\rm O32}$, where $O32=\log(\OIIIopt\lambda5007/\OII\lambda\lambda3727,3729)$; $Y$ is the adopted strong-line ratio, while $\sigma_{\rm int}$ is the intrinsic scatter of the relation, i.e. the scatter after accounting for the observational uncertainties, in dex. The $\gamma$ parameter is not fitted for the N/O and Ne/O relations, as we find no evidence for a dependence on O32 in either case; including this additional dimension does not provide any improvement in the intrinsic scatter. The C/O relation is not fitted owing to the limited sample size of only 22 objects.
}
\end{table}

Strong-line diagnostics are essential tools for deriving abundance ratios in galaxies where auroral lines are undetected, which constitutes the majority of high-redshift objects. While most previous work has focused on calibrations for oxygen abundance or N/O, in this section we also present calibrations for Ar/O, S/O and Ne/O 
using the strong-line ratios N2O2, S23/R23, Ar3O3, and Ne3O3. These ratios are chosen to minimise the impact of ICFs on the derived abundance ratios. The total abundance of an element relative to oxygen is most robustly constrained when the observed ionic ratio traces ionisation zones that overlap as closely as possible with those of $\OII$ and $\OIIIopt$; this condition is best satisfied when the IPs of the relevant ionic species are similar.

Our inferred strong-line calibrations are shown in Figure~\ref{fig:strong_lines}. Each panel displays the logarithm of an abundance ratio (N/O, Ar/O, C/O, S/O and Ne/O) as a function of the corresponding emission-line diagnostic.
The figure also shows the possible secondary dependence of these diagnostics on $\mathrm{O32} = \log(\OIIIopt\lambda5007/\OII\lambda\lambda3727,3729)$, used as a tracer of the ionisation parameter, as encoded by the colour bar on the right. We first visually inspect the distribution of each abundance diagnostic as a function of O32 to assess whether a secondary dependence is present. This inspection suggests a clear dependence for Ar/O and S/O, for which we therefore adopt a three-parameter relation of the form $\log(X/{\rm O}) = \alpha + \beta\,Y + \gamma\,{\rm O32}$, where $Y$ is the adopted strong-line ratio. 
For the other diagnostics, we retain the simpler two-parameter relations.

Across the panels, the objects with the largest deviations from the fitted relations are predominantly local galaxies with very low O32, typically $\mathrm{O32}\lesssim-0.2$. These values lie well below those sampled by most of the high-$z$ galaxies, illustrating that the local and high-$z$ datasets probe different regions of this parameter space. Where their O32 ranges overlap, however, the two samples broadly follow the same abundance--line-ratio relations. For Ar/O and S/O, including O32 accounts for much of the apparent spread, although the most weakly ionised local objects remain outliers. Whether the relations also apply to high-$z$ galaxies at such low O32 cannot be tested with the present sample: such objects are absent from it and may be particularly difficult to detect with the emission-line requirements adopted here.  No fit was performed for the relation between C/O and C3O3, given the small number of direct measurements available for this diagnostic (22 objects), and the substantial observed scatter. 

The fits are performed with \textsc{LinMix} \citep{Kelly_2007}, a Bayesian linear regression method that accounts for measurement uncertainties in both variables and provides an estimate of the intrinsic scatter.  The details of the individual diagnostics are discussed in the following sections. The best-fit parameters performed on the high-$z$ objects are reported in Table \ref{tab:bestfits}.

\subsection{N/O: the N2O2 diagnostic}

The N2O2 ratio has long been recognised as a particularly tight tracer of N/O because $\mathrm{N}^+$ and $\mathrm{O}^+$ have very similar IPs at both the lower and higher ionisation thresholds (see Table \ref{tab:IPs}), so they likely coexist within the same ionisation zone of the nebula and only have a limited dependence on its ionisation structure \citep[e.g.,][]{Florido2022}. The relation between N/O and N2O2 (Fig.~\ref{fig:strong_lines}, panel~a) is remarkably tight ($\sigma_\mathrm{int} \simeq 0.08$ dex) and displays a slope close to unity, indicating an almost linear mapping between the emission-line ratio and the abundance ratio. 

Our best fit relation is consistent within the errors with the calibration derived using the slightly smaller dataset (92 objects) presented by \citetalias{Cataldi2026} (slope = $1.04\pm 0.09$ and intercept of $-0.03 \pm 0.06$). In particular, the updated relation exhibits a moderately steeper slope, further strengthening the hint of redshift evolution that was already noted in \citetalias{Cataldi2026}. Indeed, this relation shows a clear dependence on the electron temperature: galaxies with higher $T_{\mathrm{e}}$ systematically lie toward the upper end of the relation, following a steeper trend. This difference is likely driven by the enlarged sample used in the present analysis, which includes a larger number of high-redshift, high-temperature  galaxies from the DJA. Their inclusion shifts the best-fitting relation toward a steeper slope. \citet{Cameron2026} also favours a steeper slope by studying the same relation focused on the JADES sample.


\subsection{Ar/O: the Ar3O3 diagnostic}

The Ar/O ratio, traced by Ar3O3, is shown in panel~b of Figure~\ref{fig:strong_lines}. 
For argon, we choose Ar3O3 as tracer, because $\mathrm{Ar}^{2+}$ (IP 27.6~eV) pairs with $\mathrm{O}^{2+}$ (35.1~eV), as discussed in the previous section.
The Ar3O3 diagnostic shows a clear dependence on O32, and our best-fit is in agreement with the one from \citet{Foley2026_Argon_AURORA} who reported similar coefficients. After accounting for uncertainties, the intrinsic scatter is 0.05 dex, significantly lower than the 0.14 dex obtained from a linear fit based solely on Ar3O3.

\subsection{S/O: the S23/R23 diagnostic}
For sulphur, deriving the total abundance from low-ionisation lines alone can be problematic. In particular, the $\rm S2O2 \equiv \log(\mathrm{\SII/\OII})$ diagnostic has been shown to provide a poor constraint on the S/O abundance ratio \citep[e.g.,][]{amayo_ICFs_2020, Esteban2025_alpha_elements}. This limitation is related to the ionisation structure of sulphur. The ionisation stages relevant in \hii regions and star-forming galaxies span a broad range of IPs, with S$^+$ and S$^{2+}$ producing strong emission lines in the optical, while higher ionisation stages can also contribute significantly under typical nebular conditions. In particular, S$^{2+}$ can be further ionised to S$^{3+}$, whose ionisation potential of 34.8 eV lies within the range of energies readily provided by the ionising radiation field of \hii regions. 

\citet{amayo_ICFs_2020} developed an ICF for sulphur based solely on \SII emission, but found substantial discrepancies when compared with ones obtained using both S$^+$ and S$^{2+}$.
For this reason, we adopt instead the more widely used $\rm S23/R23$ diagnostic, defined in Section \ref{subsec:direct_method}, which incorporates both low- and intermediate-ionisation species.
This choice comes at the cost of reducing the available sulphur sample, as it requires the detection of both \SII\ and \SIII\ lines. In our sample the number of objects decreases from 175 to 126. A discussion on the different sulphur diagnostics (S2O2, S3O3, and S23/R23), and their impact on the inferred S/O ratios, is presented in Appendix~\ref{app:Sulfur}. 

Panel~d of Figure~\ref{fig:strong_lines} shows the relation between S/O and S23/R23. The intrinsic scatter of this relation is 0.08 dex, based on a sample of 126 objects. The best-fit relation significantly benefits from the inclusion of the dependence on O32, which reduces the intrinsic scatter by $\sim$40\%. However, a substantial scatter is still observed, likely reflecting the complex ionisation structure of sulfur and the difficulty of accurately tracing its different ionic phases. In Appendix~\ref{app:Sulfur}, we present additional S/O calibrations based on S2O2 and S3O3 ($\equiv \log\big{(}\SIII\lambda\lambda9071,9533/ \OIIIopt\lambda5007\big{)}$). These diagnostics require fewer emission lines and may therefore be applicable to a broader range of observations, although they exhibit moderately larger intrinsic scatter than the relation presented here. We focus here on S23/R23 for consistency with the other panels, as it is the diagnostic adopted to infer S/O in our analysis.

\subsection{Ne/O: the Ne3O3 diagnostic}

Since $\mathrm{Ne}^{2+}$ and $\mathrm{O}^{2+}$ predominantly coexist in the highly ionised regions of the nebula (see discussion in Sec. \ref{subsec:Te}), Ne3O3 is a good tracer for the Ne/O ratio, as shown in panel~e of Figure~\ref{fig:strong_lines}. 
A dependence of Ne/O on O32 is also visible, particularly in the full sample. However, when restricting the analysis to the high-$z$ sample, the Ne/O relation is already relatively tight in terms of intrinsic scatter. Including O32 as an additional parameter yields only a modest improvement in the intrinsic scatter ($\sim$0.01 dex), which does not justify the introduction of a third parameter. We therefore retain the two-parameter relation for Ne/O.

\section{Observed abundance patterns}
\label{sec:abundances}

\begin{figure*}
    \centering
    \includegraphics[width=\linewidth]{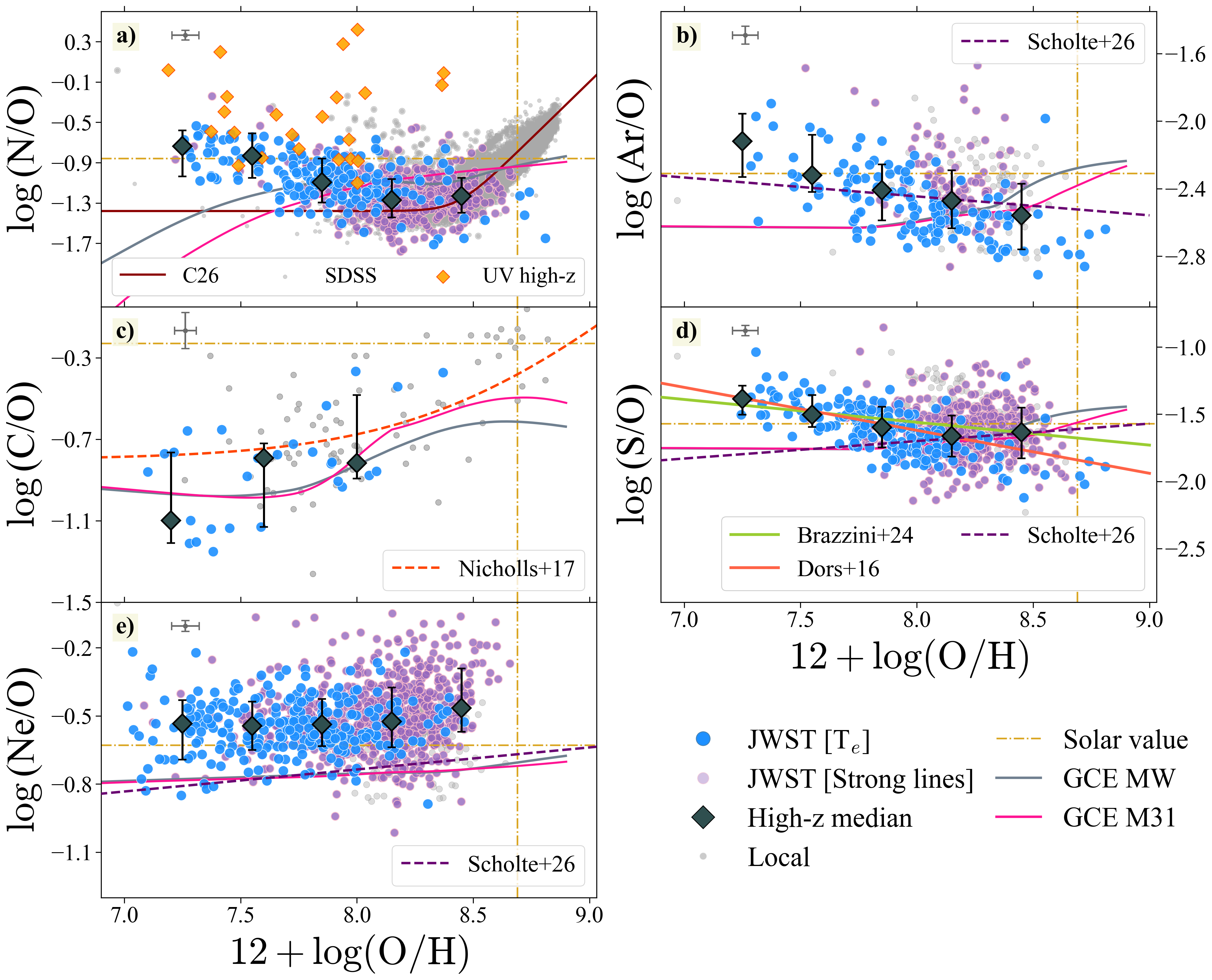}
    \caption{Abundance ratios of nitrogen (panel a), argon (b), carbon (c), sulfur (d), and neon (e) relative to oxygen, as a function of oxygen abundance. Blue circles represent the JWST high-$z$ T$_{\rm e}$ sample, comprising MARTA, JADES, and DJA galaxies, as described in Sect.~\ref{sec:MARTA} and Sect. \ref{sec:DAWN}, for which direct abundances were re-derived from the measured line fluxes, together with literature measurements from \citet{Rogers2026_CECILIA, Stanton_excels_2025, Foley2026_Argon_AURORA}. Smaller purple circles represent the JWST archival strong-line sample described in Sect.~\ref{sec:strong_line_sample}. The median uncertainties in metallicity and abundance ratio for the full high-$z$ sample are shown as grey error bars in the upper-left corner of each panel.
    Small grey circles represent local comparison galaxies, while large dark-grey diamonds indicate the median values of the combined direct-method and strong-line high-$z$ sample, computed in metallicity bins of 0.3~dex. GCE models for the Milky Way from \citet{Kobayashi2023_chemodynam} and for M31 from \citet{Arnaboldi2022, Kobayashi_Bhattacharya_M31_2023} are shown as grey and fuchsia curves, respectively. In panel~a, UV-based nitrogen abundance measurements from \citet{Ji2026} are shown as orange diamonds, the best-fitting local relation from \citetalias{Cataldi2026} is shown in dark red, and local SDSS strong-line measurements from the same study are included as small grey circles to enlarge the local comparison sample. Panel~d also includes the relations from \citet{Brazzini2024}, \citet{Dors2016}, and \citet{Scholte2026_elements}, shown in green, red, and purple, respectively. The relation from \citet{Scholte2026_elements} is also shown in panels~b and e. Finally, gold dashed lines indicate the solar abundance ratios from \citet{Asplund21}.
    }
    \label{fig:abundances}   
\end{figure*}

In this section we examine the abundance patterns as a function of oxygen abundance. We compare the high-$z$ measurements with our local benchmarks and chemical evolution models for the Milky Way from \citet{Kobayashi2023_chemodynam} and M31 from \citet{Arnaboldi2022, Kobayashi_Bhattacharya_M31_2023}. Figure~\ref{fig:abundances} provides an overview of the observed patterns. Their implications for chemical enrichment and galaxy evolution are discussed jointly in Section~\ref{sec:discussion}.

\subsection{Nitrogen}
\label{subsec:N}

Figure~\ref{fig:abundances}a confirms the enhancement in N/O at low metallicities in high-$z$ galaxies with respect to their lower-redshift counterparts previously reported by \citetalias{Cataldi2026} and \citet{Cameron2026}. Our updated compilation contains 133 direct measurements (19 of which are from \citealt{Rogers2026_CECILIA}), compared with 92 in \citetalias{Cataldi2026}, owing to the continued expansion of the DJA archive, plus 477 objects from the strong-line compilation. 
The sample spans $7.0 \lesssim 12+\log(\mathrm{O/H}) \lesssim 8.8$, extending towards lower metallicities than the minimum of approximately 7.3 reached in our previous work. The expanded dataset confirms that high-redshift galaxies tend to have higher N/O than local galaxies at low oxygen abundance, with the excess becoming more pronounced at lower metallicity.
The UV nitrogen emitters shown in Figure~\ref{fig:abundances} occupy a more extreme region, with $\log(\mathrm{N/O})\gtrsim-0.8$. This distinction is also present in the analysis of \citet{Cameron2026}, who measured N/O in 588 JADES galaxies at $z\sim1.5$--7, including 40 with direct $T_{\rm e}$-based abundances. They also find a moderate N enhancement at low metallicity. Other recent observations instead yield less pronounced differences: using deep, medium-resolution rest-optical spectra of 25 typical star-forming galaxies at $z\sim3$ from the LyC22 survey, \citet{Schaerer2026} find no significant N/O evolution within the scatter of local \hii-region measurements. 

We discuss the possible connection between these populations of N-enhanced galaxies and the additional constraints provided by carbon in Section~\ref{sec:discussion}.

\subsection{Argon}
\label{subsec:Ar}

We collect 138 galaxies with direct-method Ar/O measurements (90 from our sample and additional 48 objects from \citealt{Stanton_excels_2025, Foley2026_Argon_AURORA, Rogers2026_CECILIA}), complemented by 92 galaxies with strong-line abundances.
Panel~b of Figure~\ref{fig:abundances} shows a decreasing trend of Ar/O with increasing oxygen abundance, particularly evident in the direct-method measurements. The strong-line measurements exhibit larger scatter and suggest a somewhat flatter behaviour at higher metallicities.
A similar decrease in Ar/O with metallicity was reported locally by \citet{Esteban2025_alpha_elements}, both for star-forming galaxies and \hii regions, using abundances derived from either \ArIII\ alone or \ArIII\ together with [Ar\textsc{IV}]. Recently, \citet{Scholte2026_elements} also identified a negative correlation and derived a linear relation from approximately 50,000 star-forming galaxies at $z\sim0$ in the DESI survey \citep{DESI_2026}, with which our overall median trend is consistent within the uncertainties.

Recent high-$z$ measurements of Ar/O broadly agree with the sequence traced by our sample. The CECILIA galaxies of \citet{Rogers2026_CECILIA}, also included in our archival dataset shown in Figure ~\ref{fig:abundances}, have subsolar Ar/O on average and show no clear dependence on O/H within their sample, but they align well with our trend and extend the available constraints towards higher metallicities. The EXCELS measurements of \citet{Stanton_excels_2025} and the JADES stacks of
\citet{Isobe2026_abundance_patterns} are likewise consistent with this sequence.

\begin{figure}
    \centering
    \includegraphics[width=\linewidth]{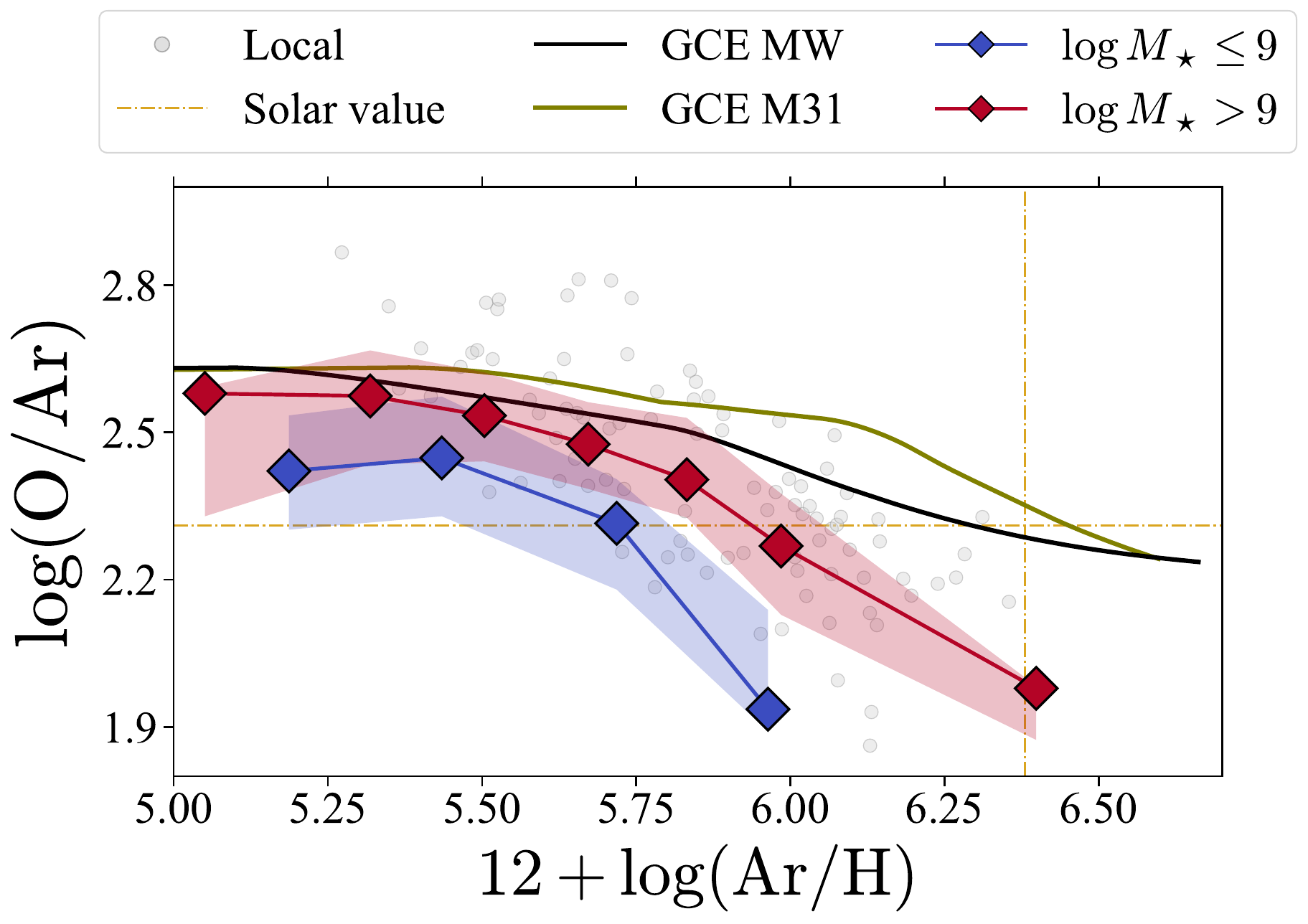}
    \caption{Abundance ratio log(O/Ar) as a function of the argon abundance 12+log(Ar/H). The blue and red diamonds represent, respectively, the median values for the log(M$_*$/M$_{\odot}$) $\lesssim$ 9 and log(M$_*$/M$_{\odot}$) > 9 subsamples. Median values are computed in adaptive bins in Ar/H, requiring at least five objects per bin. 
    The black and green solid curves show chemical evolution models for the Milky Way from \citet{Kobayashi2023_chemodynam} and for M31 from \citet{Arnaboldi2022, Kobayashi_Bhattacharya_M31_2023}, respectively, while the yellow dotted lines indicate the solar abundance ratios from \citet{Asplund21}.}
    \label{fig:OAr_ArH}   
\end{figure}

In Fig.~\ref{fig:OAr_ArH}, we examine the distribution of galaxies in the O/Ar--Ar/H plane and its dependence on stellar mass. Following the work of \citet{Bhattacharya2026}, we divide the sample into two stellar-mass bins, $\log(M_*/M_\odot)\leq9$ and $\log(M_*/M_\odot)>9$, and compute the median O/Ar in adaptive bins of Ar/H, each containing at least five galaxies. The shaded regions indicate the 16th--84th percentile range of the observed O/Ar distribution within each bin.
Despite the limited sample size, a clear mass-dependent separation emerges. At fixed Ar/H, the median sequence of higher-mass galaxies lies systematically $\sim0.1$--$0.2$~dex above that of lower-mass systems. The higher-mass sequence broadly follows the GCE tracks shown in the figure, although its last two bins at the highest Ar/H exhibit a downturn towards lower O/Ar. Lower-mass galaxies, spanning approximately $7.5\lesssim\log(M_*/M_\odot)\lesssim9$, occupy systematically lower O/Ar values over the overlapping Ar/H range.

Possible interpretations of the Ar/O behaviour and its mass dependence in the O/Ar--Ar/H plane are discussed in more detail in Section~\ref{sec:discussion}.

\subsection{Carbon}
\label{subsec:C}

We measure C/O in 22 galaxies, restricted to systems at $z\gtrsim3$ for which the relevant rest-frame UV carbon lines fall within the observed spectral range. Because \OIII$\lambda1666$ is generally unavailable, we infer C/O using \CIII$\lambda\lambda1907,1909$ and \OIIIopt$\lambda5007$. The large wavelength separation makes the \CIII/\OIIIopt\ ratio sensitive to
differential dust attenuation. However, most galaxies in the sample have low $E(B-V)$, which limits the impact of the assumed attenuation curve. The
largest value, $E(B-V)=0.45$, corresponds to a correction of approximately a factor of seven in \CIII/\OIIIopt\ relative to the uncorrected ratio. This is a substantial correction, but only $\sim$5\% of the galaxies have $E(B-V)>0.3$, so corrections of this size affect a small fraction of the sample.

Fig.~\ref{fig:abundances}c suggests an increase of C/O with oxygen abundance. This behaviour resembles that observed in local star-forming galaxies and \hii\ regions \citep{Garnett1995_C/O, Kobulnicky1998_CNO, Esteban_2002_RL_abundances, esteban_keck_2009, berg_carbon_2016, nicholls_abundance_2017, Berg_CNO_2019}. The trend is also broadly consistent with the chemical evolution models for the Milky Way and M31 shown in the figure, and the high-z values lie within the uncertainties of the local C/O--O/H relation of \citet{nicholls_abundance_2017}. However, the limited sample size prevents firm conclusions. The implications of this result, particularly in combination with the elevated N/O, are discussed in
Sect.~\ref{sec:discussion}.

\subsection{Sulfur}
\label{subsec:S}

We collect 143 direct S/O measurements (126 from our sample and 17 from \citealt{Rogers2026_CECILIA}) in the high-$z$ sample, complemented by 227 galaxies with strong-line abundances. Panel~d of Figure~\ref{fig:abundances} shows a decreasing S/O trend with increasing oxygen abundance, broadly similar to that observed for Ar/O. The strong-line measurements follow a similar overall trend, but suggest a flattening at higher metallicities. These measurements dominate the sample above $12+\log(\mathrm{O/H})\approx 8.1$, as expected given that they are not subject to the detectability limitations imposed by weak auroral lines. The median S/O values are consistent within the uncertainties with the decreasing linear relations derived for local galaxies by \citet{Dors2016} and \citet{Brazzini2024}, except for the highest-metallicity bin, which is dominated by strong-line measurements and lies above both relations. In this regime, the median S/O values tend to flatten, approaching the relations predicted by local GCE models and the empirical relation derived by \citet{Scholte2026_elements}. This flattening could be related to dust depletion, which becomes increasingly important in metal-rich environments. If oxygen is preferentially depleted onto dust grains relative to sulphur, the measured gas-phase S/O ratio would increase, potentially counteracting the declining trend observed at lower metallicities.

The local literature presents a mixed picture. Decreasing relations have been reported in several studies \citep{1988MNRAS.231..257V, Dors2016, Diaz_Zamora_2022, Brazzini2024, Perez-Diaz_2024}, whereas most find approximately constant or increasing S/O \citep{Garnett1997, kennicutt_composition_2003, izotov_chemical_2006, berg_chaos_2020, Esteban2025_alpha_elements, Scholte2026_elements}. These differences highlight the importance of the adopted samples and abundance prescriptions.
At high redshift, our measurements overlap well with those from CECILIA presented by \citet{Rogers2026_CECILIA}. Their sample, of approximately 20 galaxies, does not establish a metallicity dependence on its own, but extends the available direct measurements to $12+\log(\mathrm{O/H})\sim8.8$. 

 \subsection{Neon}
\label{subsec:Ne}

Ne/O is the best-sampled abundance ratio in our analysis, with 188 direct measurements (180 from our sample and eight from \citealt{Stanton_excels_2025}), enabled by the widespread detection of \NeIII$\lambda3869$ in JWST spectra, complemented by 395 strong-line measurements. As shown in panel~e of Figure~\ref{fig:abundances}, the two samples occupy broadly similar regions of the Ne/O--O/H plane, although the strong-line measurements exhibit larger scatter. The combined median Ne/O remains approximately constant up to $12+\log(\mathrm{O/H})\sim8.2$, with an intrinsic scatter of $\sim0.16$~dex, and shows no significant redshift dependence over $z\sim1$--7. At higher metallicities, where the strong-line sample extends the coverage to $12+\log(\mathrm{O/H})\sim8.6$, the median values suggest a mild increase of $\lesssim0.1$~dex.

The nearly constant Ne/O at lower metallicities is consistent with the EXCELS measurements of \citet{Stanton_excels_2025} and the JADES stacks of \citet{Isobe2026_abundance_patterns}. A weak increase towards higher O/H has also been reported in local studies, including \citet{izotov_chemical_2006}, who find a variation of approximately 0.1~dex, as well as \citet{Esteban2025_alpha_elements} and \citet{Scholte2026_elements}. Similarly, the local relation of \citet{Scholte2026_elements} and the chemical evolution models shown in Figure~\ref{fig:abundances} predict only a modest variation across the metallicity range probed by our sample. The slight increase suggested by our strong-line measurements above $12+\log(\mathrm{O/H})\sim8.2$ is therefore qualitatively consistent with these expectations, although both the local relation and the models lie approximately 0.2~dex below our median measurements.

Part of the offset relative to the local relation may arise from differences in the adopted ionisation correction factors (ICFs). While \citet{Scholte2026_elements} employ the prescription of \citet{izotov_chemical_2006}, we adopt that of \citet{amayo_ICFs_2020}. As discussed in Section~\ref{subsec:ICFs}, applying the former prescription to our sample would lower the inferred Ne/O by approximately 0.1~dex, reducing the discrepancy. The observed distributions nevertheless show partial overlap.
The increase in Ne/O at higher metallicities could also be related to dust depletion. Since neon is not depleted onto dust grains, whereas oxygen can be, an increasing oxygen depletion at higher metallicities would lead to an apparent enhancement of the gas-phase Ne/O ratio. Interestingly, this increase occurs over approximately the same metallicity range where the S/O relation begins to flatten, suggesting that dust depletion could contribute to both features. 

\section{Discussion}
\label{sec:discussion}

The abundance patterns presented above provide complementary constraints on the enrichment histories of high-redshift galaxies. N/O is moderately enhanced at low oxygen abundance, while C/O remains broadly consistent with the local metallicity-dependent relation. S/O and Ar/O decrease with increasing O/H, and Ne/O shows no significant dependence on metallicity or redshift. Interpreting these observations jointly can help distinguish changes in stellar enrichment from the effects of gas flows and abundance methodology.
A limitation is that the different elements are not measured in identical subsamples. In particular, the C/O sample is smaller and predominantly probes higher redshifts than the rest-optical measurements. The following discussion distinguishes the implications of the overall distributions from those of simultaneous abundance measurements in individual galaxies.

\subsection{Constraints from nitrogen and carbon}
\label{subsec:discussion_CN}

JWST observations have identified galaxies with strongly enhanced nitrogen abundances, often selected through \NIV$\lambda1486$ and \NIII$\lambda1750$ emission \citep[e.g.,][]{Morel_2026_nitrogen, Ji2026}.
These extreme UV N-emitters must be distinguished from the moderate N/O enhancement found through rest-optical diagnostics in our sample and by \citet{Cameron2026}. The distinction does not necessarily imply unrelated enrichment mechanisms, but differences in the strength, spatial extent, and duration of enrichment may be important.

Stacking analyses suggest that nitrogen enrichment may also extend beyond the rare, most extreme emitters. \citet{2025ApJ...982...14H}, using approximately 1000 low-resolution PRISM spectra, report an increase in UV-based N/O at $z>4$. They argue that enrichment mechanisms invoked for individual N-emitters could operate more weakly in the wider star-forming population. 

Several mechanisms have been proposed for nitrogen enhancement, including WR, very massive, and supermassive stars \citep{charbonnel_gnz11_2023, Nandal_2024_N_enhancement_popIII, kobayashi_ferrara_gnz11_2024, marques-chaves_Nitrogen_2024, watanabe_2024}, enrichment in dense young clusters associated with proto-globular cluster formation \citep{2023A&A...680L..19D, 2025MNRAS.544..483G, Berg2026, Ji2026}, bursty star-formation histories \citep{Bhattacharya2026_N_emitters_DESI, McClymont2026}, preferential loss of oxygen-rich ejecta \citep{2025A&A...697A..96R}, and dilution by metal-poor gas \citep{2025ApJ...981..136S}.
We refer to \citetalias{Cataldi2026} for a detailed discussion of these possibilities. Here, we focus on how carbon can provide additional constraints.

Subsolar C/O is common among metal-poor galaxies. High-$z$ measurements generally remain compatible with the local C/O--O/H relation \citep{arellano_cordova_2020_abundances, isobe_CNO_2023, jones_CO_z6_2023,
Castellano_2024_GHz12, topping_z6_lens_2024, Morel_2026_nitrogen}.
For example, \citet{arellano_cordova_2024_CLASSY} measured C/O at $z\sim8.5$ using the \CIII\ doublet and found a value consistent with the scatter of lower-redshift metal-poor dwarfs. Similarly, \citet{topping_z6_lens_2024} measured subsolar C/O in RXCJ2248-ID, compatible with enrichment dominated by low-metallicity massive stars.
The increase of C/O towards higher O/H can reflect both metallicity-dependent carbon yields from massive stars and delayed enrichment by lower-mass stars. Stronger stellar winds at higher metallicity can enhance the carbon contribution from massive stars, while LIMS return carbon on longer timescales \citep{Akerman_2004_CO, Carigi_2005_CNO_problems}.
These contributions help explain the local increase in C/O, particularly above $12+\log(\mathrm{O/H})\sim8.0$ \citep{nicholls_abundance_2017}.

Against this background, elevated N/O combined with approximately normal C/O at fixed O/H would indicate nitrogen enrichment without a comparable carbon enhancement. A possible explanation involves the different yields and lifetimes of AGB stars: lower-mass AGB stars, with initial masses of $\sim 2$--$4\,M_\odot$, can release the carbon previously brought to their envelopes through third dredge-up \citep{Karakas_2010_AGB, Karakas_Lugaro_2016_AGB}; more massive AGB stars can instead undergo HBB, which converts carbon into nitrogen through the CNO cycle \citep{Karakas_2010_AGB, Ventura_DAntona_2009}.
Their shorter lifetimes allow this nitrogen-enriched material to be returned to the ISM before the carbon contribution from lower-mass AGB stars becomes important. The relevant mass ranges and yields depend on metallicity and stellar-evolution prescriptions. This scenario does not require an episodic SFH, although the duration and shape of star formation determine the relative contributions of the different stellar populations.
The EXCELS observations of \citet{2025MNRAS.540.2991A} illustrate this possibility. Their two galaxies at $z\simeq5$ have elevated N/O and relatively low C/O, with carbon abundances consistent with CCSN yields. Comparison with the models of \citet{kobayashi_isotopes_MW_2011} favours a bottom-heavy IMF in their analysis, but the uncertainties also permit a standard IMF.

Rapid enrichment by H-burning products provides another explanation, particularly for extreme, young N-emitters. Several systems have C/N and N/O approaching the abundance patterns expected for CNO-processed material
\citep{isobe_CNO_2023, Castellano_2024_GHz12, Senchyna_2024_GNz11}.
Potential sources include WN stars, very massive or supermassive stars, and tidal disruption events. By contrast, a substantial contribution from carbon-rich He-burning ejecta would also affect C/O.
The observations of \citet{Berg2026} provide direct evidence for WR stars in RXCJ2248-ID3, where enhanced N/O and low C/O are interpreted as signatures of WN-dominated enrichment. This establishes the relevance of WR enrichment in at least some systems, but does not imply that it accounts for the moderate enhancement in the wider star-forming population. The duration of the nitrogen-rich phase and the dilution of stellar ejecta into the surrounding gas are essential parts of this interpretation.
The compilation of 45 N-emitters by \citet{Morel_2026_nitrogen} likewise shows elevated N/C, while C/O is generally comparable to local values at similar O/H. This favours enrichment by H-burning products without a corresponding systematic enhancement from carbon-rich He-burning material. The increasing incidence of N-emitters towards higher redshift further motivates enrichment associated with dense young stellar clusters, potentially containing WR, VMSs, SMSs.

Gas flows can modify these patterns, but different processes have distinct effects. In the limiting case of dilution by gas containing negligible metals, O/H decreases while N/O and C/O remain unchanged. A galaxy can therefore appear displaced from the local abundance-ratio relations without newly producing nitrogen. Conversely, preferential oxygen loss, with carbon and nitrogen retained, raises both N/O and C/O while leaving N/C unchanged. Elevated N/C without a comparable C/O enhancement would therefore be difficult to explain through oxygen loss alone. These simplified predictions do not exclude combinations of gas flows and stellar enrichment, but clarify why simultaneous C, N, and O measurements are valuable.

\subsection{Sulfur and argon in the context of alpha-element enrichment}
\label{subsec:discussion_alpha}

The similar behaviour of S/O and Ar/O motivates a joint discussion of these elements, given their related nucleosynthetic origins \citep{kobayashi_origin_2020}. Our sample exhibits decreasing Ar/O with increasing O/H, broadly consistent with the trends reported by \citet{Esteban2025_alpha_elements, Scholte2026_elements}. Although some high-redshift studies do not identify a significant metallicity dependence within their individual samples, their measurements broadly overlap with the sequence traced here \citep{Stanton_excels_2025, Rogers2026_CECILIA,
Foley2026_Argon_AURORA}. The relatively high Ar/O values at low oxygen abundance raise the question of how a substantial argon contribution can develop while the gas remains metal-poor.

Argon is produced by both CCSNe and SNe~Ia, whereas oxygen is predominantly supplied by massive stars \citep{Iwamoto1999, kobayashi_origin_2020}. The balance between these enrichment channels can therefore modify Ar/O. \citet{Bhattacharya2025_OAr} explored an illustrative model involving intermittent star formation and substantial pristine gas accretion to explain several high-redshift galaxies with low O/Ar. Following an initial burst, continued enrichment by SNe~Ia during a period of reduced SF can increase Ar/O (and hence lower O/Ar). Subsequent pristine gas accretion can lower O/H while initially preserving the abundance ratio Ar/O, allowing relatively Ar-rich gas to occur at low metallicity. Renewed SF from a second burst then supplies fresh oxygen, producing a loop in the abundance plane and potentially bringing the system back towards the standard enrichment sequence. Such histories can explain the positions of individual galaxies.

Additional information is however provided by the mass dependence observed in Figure~\ref{fig:OAr_ArH}: at a given Ar/H, lower-mass galaxies have systematically lower O/Ar. The statistical analysis presented above supports this association, suggesting that the distribution is not dominated by displacements of individual systems.
A similar mass-dependent trend is identified in local SDSS galaxies by \citet{Bhattacharya2026}, who interpret it in terms of star-formation efficiency. In their framework, more massive galaxies have more efficient SF, 
so enrich their gas in oxygen rapidly through CCSNe, reaching higher metallicities before the delayed contribution from SNe~Ia becomes substantial. Less efficient, more prolonged SF in lower-mass galaxies allows SNe~Ia to contribute while Ar/H is still relatively low, resulting in lower O/Ar at a given Ar/H. Our observed mass dependence is qualitatively consistent with this interpretation. Importantly, \citet{Bhattacharya2026} find that the mass-dependence becomes unclear when including galaxies that have undergone perturbations, such as metal-poor gas accretion or outflows. The persistence of a mass-dependent sequence in our sample -- and its statistical significance --  therefore suggests that large accretion-driven excursions do not dominate the population, although they could of course affect individual objects \citep{Kobayashi2023_chemodynam, Bhattacharya2026}. 

Slower chemical enrichment, potentially sustained over an extended period, thus provides a plausible interpretation of the relatively high Ar/O values in our lower-mass galaxies. However, the correspondence between these mass-dependent sequences and the overall Ar/O--O/H trend remains to be established quantitatively. The Milky Way and M31 tracks provide illustrative comparisons, while models tailored to the masses, redshifts, and physical conditions of our sample are required to test whether this framework can simultaneously reproduce all the observed distributions.

Dust depletion provides an additional complication for sulfur. Oxygen is likely more strongly depleted than sulfur, so that increasing depletion at higher metallicity would raise gas-phase S/O rather than produce the decrease observed here \citep{jenkins_depletion_2009}. More efficient dust formation at higher metallicity could strengthen this effect \citep{Galliano2021_dust}.
Under this picture, oxygen depletion is unlikely to explain the observed S/O trend. However, the flattening suggested by the strong-line measurements at $12+\log(\mathrm{O/H})\gtrsim8.0$ could be partly driven by increasing oxygen depletion, which would counteract the declining S/O trend. This could also help explain why the median values at higher metallicities approach the approximately constant relations predicted by local chemical evolution models. 
Nevertheless, uncertainties in sulfur depletion, ionisation corrections, and temperature estimates prevent us from identifying nucleosynthesis as the sole cause.

Finally, Ne/O offers a useful comparison. Oxygen and neon are predominantly produced by massive stars, with comparatively small contributions from AGB stars and SNe~Ia, so their ratio is expected to be less sensitive to delayed enrichment \citep{izotov_chemical_2006, kobayashi_origin_2020}. Its approximately constant behaviour up to $12+\log(\mathrm{O/H})\sim8.2$ is consistent with this expectation. At higher metallicities, however, the strong-line measurements suggest a mild increase in Ne/O, occurring over a similar metallicity range to the flattening observed in S/O. Both features could be related to increasing oxygen depletion onto dust grains, which would raise the measured gas-phase abundance ratios. This interpretation has also been proposed to explain the mild increase in Ne/O reported in some local samples, since neon is not incorporated into dust grains \citep[e.g.,][]{izotov_chemical_2006}. However, uncertainties in the ionisation corrections may also contribute to the observed behaviour. In particular, \citet{Esteban2025_alpha_elements} highlight limitations of Ne ICFs at low ionisation, as predicting the relative ionisation structures of Ne and O depends on the adopted ionising spectra and treatment of stellar opacities and winds \citep{Smith_Norris_2002, Perez-Montero2007_Ne_Ar}.

In summary, the weak overall metallicity dependence of Ne/O, with a possible mild increase at higher O/H, and the absence of significant redshift evolution support the use of neon-based strong-line diagnostics. The bright \NeIII$\lambda3869$ line remains accessible over a wide redshift range, making such diagnostics very useful for JWST observations \citep{Perez-Montero2007_Ne_Ar, sanders_ceers_2023, 10.1093/mnras/staf834}.

\subsection{Selection effects in the abundance planes}
\label{sec:selection_results}

In this section, we aim to answer the following question: if high-redshift galaxies obey all the currently established high-redshift scaling relations, but their relative abundances follow the local abundance-metallicity relations, would they be detectable with typical integrations of our archival JWST/NIRSpec data? We simulate a population of `regular' high-z galaxies, as described in detail in Appendix \ref{app:model}, following a forward-modelling approach. We considered a galaxy as `detectable' for a specific abundance measurement if both the auroral line \OIIIopt$\lambda$4363 and the emission line associated with the abundance diagnostic under consideration (e.g. \ArIII\ for Ar/O) are detected above the adopted signal-to-noise threshold of 3. The other nebular lines entering the abundance diagnostics (e.g. \OIIIopt$\lambda$5007) are always brighter, so they do not drive the detectability. We focus this comparison on galaxies with direct-method abundances, since these drive the observed trends at low metallicity. In particular, we aim to assess whether the detectability requirement on \OIIIopt$\lambda$4363 could introduce selection effects capable of producing or enhancing these trends.

\begin{figure*}[]
    \centering
    \sidecaption
    \includegraphics[width=12cm]{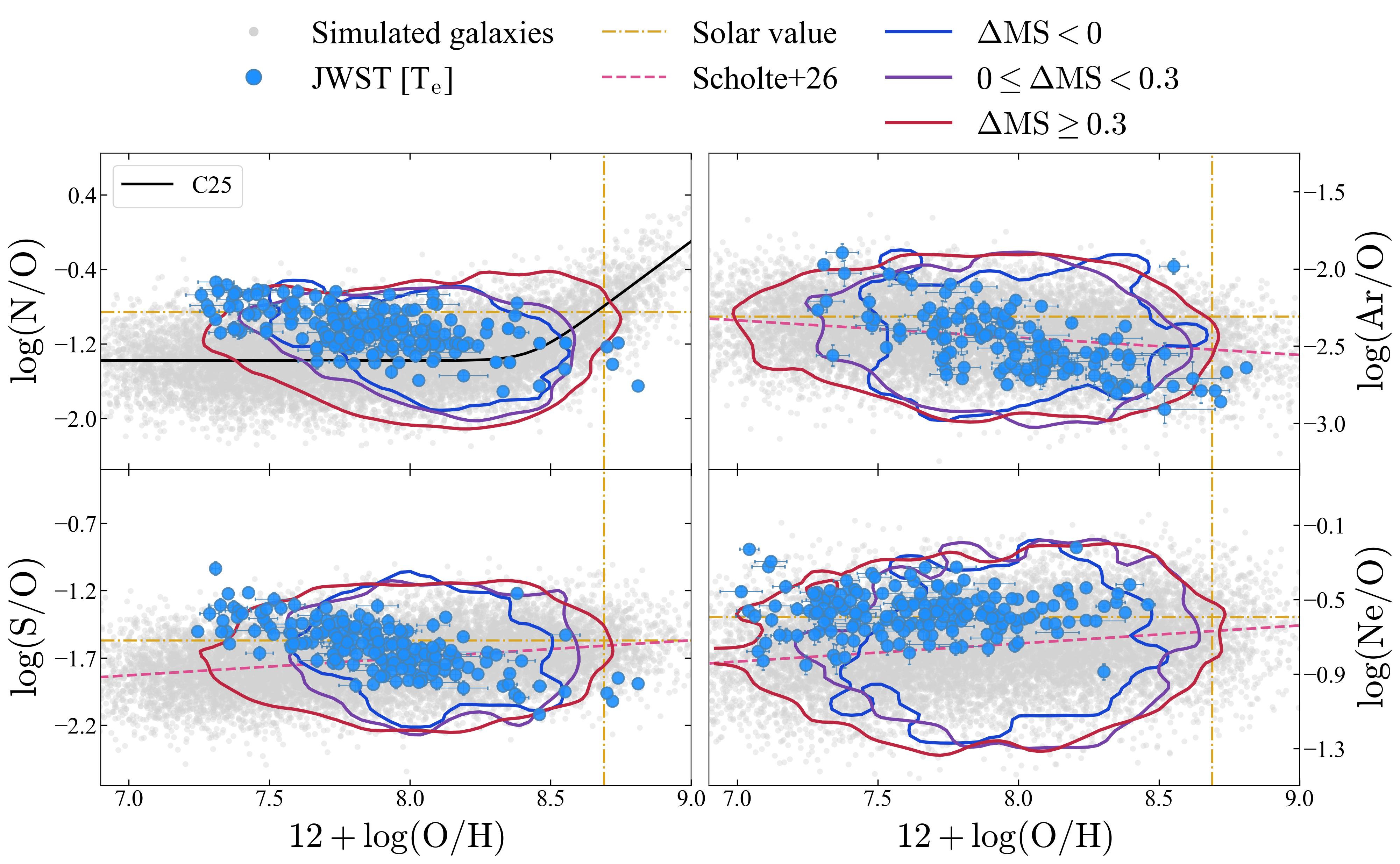}
    \caption{Four abundance ratios as a function of metallicity: N/O (upper left), S/O (upper right), Ar/O (lower left), and Ne/O (lower right). Galaxies from the JWST T$_{\rm e}$ sample are shown as blue circles, while the population generated by the forward model is represented by small grey circles. The contours enclose the 97\% highest-density regions of the simulated galaxies expected to be detectable in a 5-hour JWST/NIRSpec observation, requiring $\mathrm{S/N}>3$ in both the relevant emission line for each abundance diagnostic (\NII, \ArIII, \SII, or \NeIII) and the \OIIIopt auroral line. This exposure time is representative of the median integration time of the observing programmes currently included in the DJA. Different contour colours correspond to different simulated $\Delta$MS bins: $\Delta\mathrm{MS}<0$ dex (blue), $0\leq\Delta\mathrm{MS}\leq0.3$ dex (purple), and $\Delta\mathrm{MS}>0.3$ dex (red).}
    \label{fig:detectability}
\end{figure*}

The results are presented in Figure~\ref{fig:detectability}, which reproduces four of the abundance--metallicity planes shown in Figure~\ref{fig:abundances}, together with the simulated galaxy population. The contours enclose the 97\% highest-density regions of the simulated galaxies satisfying the adopted detectability criteria for a 5-hour JWST/NIRSpec observation, representative of the median exposure time of the observing programmes currently included in the DJA. Different colours correspond to different $\Delta$MS bins.

Several conclusions can be drawn from this figure. Most observed galaxies fall within the regions occupied by simulated galaxies that satisfy the adopted detection criteria; only a few lie outside the plotted contours. These cases are recovered when the contours are extended beyond the adopted 97\% enclosed fraction towards the full detectable distribution; moreover, some of these galaxies were observed for longer than the representative 5-hour exposure assumed here, and would be detectable at correspondingly lower flux limits.
Detectability depends strongly on a galaxy's position relative to the star-forming main sequence. Simulated galaxies with $\Delta\mathrm{MS}<0.3$ are detectable down to $12+\log(\mathrm{O/H})\simeq7.8$--8.0, but are largely absent from the detectable population at lower metallicities. Below this value, the observed galaxies can be reproduced primarily by simulations in the $\Delta\mathrm{MS}\geq0.3$ bin, which extend down to $12+\log(\mathrm{O/H})\sim7.2$. Thus, with $\sim$5-hour JWST/NIRSpec exposures, direct-abundance samples at the lowest metallicities preferentially select galaxies with SFRs elevated above the main sequence.

Within the assumptions of the forward model, this selection does not appear to produce the abundance--metallicity trends by itself in all cases. Although the detection criteria restrict the metallicity range accessible in each $\Delta\mathrm{MS}$ bin, the detectable simulated galaxies broadly follow the abundance relations assigned to the parent population. In the Ne/O panel, the local relation of \citet{Scholte2026_elements}, used to generate the simulated population, lies in median below the observed high-z measurements. This offset possibly reflects differences in the adopted ICFs, since \citet{Scholte2026_elements} adopt the prescriptions of \citet{izotov_chemical_2006}, whereas we use  \citet{amayo_ICFs_2020}, which yield median offsets of $\sim0.1$ dex from the former, as previously pointed out in Section  \ref{subsec:ICFs}. Nevertheless, the observed Ne/O values remain consistent with the simulated distribution within its scatter, and neither the observed nor the detectable simulated sample shows a clear Ne/O trend with metallicity. The detection cut mainly determines which part of that relation is accessible, with the lowest-metallicity galaxies appearing almost exclusively in the highest-$\Delta\mathrm{MS}$ bin. A similar result is found for Ar/O: the simulated and observed populations show little discrepancy, and nearly all galaxies fall within the detectable region of the parameter space.

The N/O plane presents a different picture. At low metallicity, the detectable simulated galaxies preferentially have enhanced N/O, producing a distribution consistent with the trend observed in the direct-method sample. This suggests that selection effects could account for part of the N/O enhancement at low O/H, even if the underlying population follows the local abundance--metallicity relation. A weaker similar preference towards higher abundance ratios at low O/H is also visible in the S/O plane, hinting that selection may contribute to its observed trend. 

Overall, these results suggest that detectability shapes the composition of the observed sample, particularly at low metallicity, but does not fully account for the observed abundance patterns of Ne/O and Ar/O, while potentially contributing to the trends observed for N/O and, to a lesser extent, S/O. The comparison remains qualitative because it adopts a single representative 5-hour exposure, while the observations span a range of integration times. 

Another noteworthy feature is that essentially no galaxies with super-solar metallicities are predicted to be detectable, for each considered abundance ratio and $\Delta$MS bin. This limitation is not driven by the individual diagnostic lines, but rather by the requirement of detecting the temperature-sensitive \OIIIopt$\lambda$4363 auroral line. As illustrated in Appendix \ref{app:model}, the \OIIIopt$\lambda$4363/\OIIIopt$\lambda$5007 ratio decreases steeply with increasing metallicity. More efficient cooling in metal-rich nebulae lowers the electron temperature, weakening the \OIIIopt$\lambda$4363 auroral line until it becomes difficult to detect with JWST/NIRSpec. Its measurement can also be complicated by the nearby [Fe\,II]~$\lambda$4360 emission line, whose contribution becomes particularly important when \OIIIopt~$\lambda$4363 is faint \citep{curti_new_2017, Shapley2025_z7}. These limitations make direct-method abundances based on \OIIIopt~$\lambda$4363 particularly challenging at high metallicity. Other auroral lines, however, do not decline as steeply: the \OII~$\lambda\lambda$7320,7330 doublet may remain measurable in more metal-rich systems, offering an alternative route to constraining the electron temperature.

\section{Conclusions}
\label{sec:conclusions}

We have presented a systematic study of the abundance patterns of C, N, Ne, S, and Ar in star-forming galaxies at $z\sim1$--7, combining ultra-deep MARTA observations with archival JWST/NIRSpec spectroscopy. Our direct-method sample, based on detections of the auroral \OIIIopt$\lambda4363$ line, spans $7\lesssim 12+\log(\mathrm{O/H})\lesssim8.5$, allowing us to investigate multiple abundance ratios tracing different nucleosynthetic channels and enrichment timescales. Our main findings are summarised below:

\begin{enumerate}

\item In Section \ref{sec:calibrations}, we derived new empirical strong-line calibrations for N/O, Ne/O, Ar/O, and S/O, anchored to the high-redshift direct-method sample. N2O2 and Ne3O3 provide particularly tight relations, with intrinsic scatters of $\sim0.07$ and $\sim0.04$ dex, respectively, while Ar/O and S/O show a significant secondary dependence on O32, which we include in their calibrations. These relations extend the abundance analysis to larger samples and higher metallicities, where auroral lines become difficult to detect. The limited C/O sample prevents us from deriving a robust calibration.

\item In Section \ref{subsec:N}, we confirm the systematic N/O enhancement of high-redshift galaxies relative to local systems, particularly at low metallicities, previously reported in \citetalias{Cataldi2026}. The strong-line measurements support this result and extend the comparison to a larger sample and wider metallicity range.

\item In Section \ref{subsec:Ar}, direct-method measurements reveal decreasing Ar/O with increasing oxygen abundance, broadly supported by strong-line measurements, which suggest a possible flattening at higher metallicities. We also identify a stellar-mass dependence in the O/Ar--Ar/H plane, with lower-mass galaxies exhibiting lower O/Ar at fixed Ar/H. This is qualitatively consistent with less efficient star formation and a larger relative contribution from delayed SNe Ia enrichment in lower-mass systems.

\item In Section \ref{subsec:C}, the available C/O measurements tentatively suggest an increase with oxygen abundance, broadly consistent with the local relation and chemical-evolution models. This may reflect increasing contributions from delayed carbon production by intermediate-mass stars and AGB winds, although the limited sample prevents firm conclusions.

\item In Section \ref{subsec:S}, S/O shows a decreasing trend with metallicity, similar to Ar/O. Strong-line measurements suggest a flattening above $12+\log(\mathrm{O/H})\approx8.1$, where median values approach local chemical-evolution predictions. Although oxygen depletion onto dust cannot explain the overall decline, it may contribute to the flattening at higher metallicities by increasing gas-phase S/O.

\item In Section \ref{subsec:Ne}, Ne/O remains approximately constant up to $12+\log(\mathrm{O/H})\sim8.2$, with an intrinsic scatter of $\sim0.16$ dex and no significant redshift evolution over $z\sim1$--7. Strong-line measurements extend to $12+\log(\mathrm{O/H})\sim8.6$ and suggest a mild increase ($\lesssim0.1$ dex) at higher metallicities, potentially related to oxygen depletion onto dust. The relative stability of Ne/O supports the use of neon-based diagnostics at high redshift.

\item In Section \ref{sec:discussion}, we investigate observational selection effects through a forward model (Appendix \ref{app:model}). The requirement of detecting \OIIIopt$\lambda4363$ and the relevant abundance-sensitive lines limits the accessible metallicity range, particularly below the star-forming main sequence. While selection primarily restricts the observable range for Ne/O and Ar/O, it preferentially selects galaxies with enhanced N/O at low metallicity, producing a distribution consistent with observations. A weaker effect is found for S/O, suggesting that selection may contribute to these abundance trends alongside intrinsic chemical enrichment.

\end{enumerate}

Taken together these findings reveal distinct abundance--metallicity behaviours among elements with different nucleosynthetic origins. The S/O and Ar/O trends and the stellar-mass dependence of O/Ar provide new constraints for chemical-evolution models; our forward-modelling highlights the need to account for observational selection when interpreting abundance patterns, especially for N/O at low metallicity. Larger samples with homogeneous sensitivity and GCE models tailored to high-$z$ galaxies will be essential to disentangle nucleosynthetic processes, enrichment timescales, and observational biases.


\begin{acknowledgements}
FB, FM, AM, MG, IL, CM, AF, EB, GC acknowledge support from the INAF Fundamental Astrophysics programme 2022, 2023 and 2024. 
IL, FB, and AM acknowledge support from PRIN-MUR project ``PROMETEUS'' financed by the European Union -- Next Generation EU,
Mission 4 Component 1 CUP B53D23004750006 and C53D2300080006. 
GC acknowledges the COSMOWEBB project “A new era for cosmology: exploiting the JWST revolution”, funded by the Italian National Institute for Astrophysics (INAF) through the PRORIS programme (PROgramma di RIcerca Spaziale di base), CUP C83C26000300005.
GV acknowledges financial support from the Italian National Institute for Astrophysics (INAF) under the IAF - Astrophysics Fellowships in Italy grant CUP C59J21034720001 - ``AD MAJORA''.
Some of the data products presented herein were retrieved from the Dawn JWST Archive (DJA). DJA is an initiative of the Cosmic Dawn Center (DAWN), which is funded by the Danish National Research Foundation under grant DNRF140.
\end{acknowledgements}

\bibliography{model}{}
\bibliographystyle{aa}

\begin{appendix}

\section{JWST archival data sources}
\label{app:jwst_programs}

\begingroup
\centering
\captionof{table}{JWST/NIRSpec programs used in this work.}
\label{tab:jwst_programs}

\begin{tabular}{llc}
\hline\hline
Survey / PI & Program ID & Reference \\
\hline
JADES              & 1180, 1181,1210, 1286, 1287, 3215  & \citet{DEugenio2024_JADES_DR3} \\
GTO (M. Stiavelli)  & 1199  & \citet{Stiavelli2023_GTO} \\
GTO (G. Rieke)      & 1207  & --- \\
GTO WIDE            & 1212, 1215  & \citet{Maseda_WIDE_2024} \\
GTO (C. Willott)    & 1222  & \citet{Christensen2023} \\
GLASS-ERS           & 1324  & \citet{Mascia2024_GLASS} \\
CEERS-ERS           & 1345  & \citet{Finkelstein_ceers_paper1_2023} \\
(B. Maseda)         & 1671  & \citet{maseda_xmps_lae_2023} \\
Bluejay             & 1810  & \citet{Belli2024_bluejay} \\
LyC-22 (D. Schaerer) & 1869  & --- \\
AURORA              & 1914  & \citet{Shapley2025_AURORA} \\
SUSPENSE            & 2110  & \citet{Slob2024_SUSPENSE} \\
(D. Stark)          & 2478  & \citet{2025ApJ...980..225T} \\
CECILIA             & 2593  & \citet{Strom2023CECILIA:} \\
SMACS-0723 ERO       & 2736  & \citet{pontoppidan_ERO_2022} \\
GTO (M. Stiavelli)  & 2758  & --- \\
JADES Ultra-Deep     & 3215  & \citet{Eisenstein_JADES_2023} \\
ASPIRE (F. Wang)     & 3325  & --- \\
EXCELS              & 3543  & \citet{2024MNRAS.534..325C} \\
DEEPDIVE            & 3567  & \citet{Ito2026_DEEPDIVE} \\
(E. Nelson)          & 4106  & --- \\
RUBIES              & 4233  & \citet{de_graaff_25} \\
(J. Antwi-Danso)     & 4318  & --- \\
(S. Fujimoto)        & 4762  & --- \\
GLIMPSE (S. Fujimoto) & 9223  & --- \\
\hline
\end{tabular}
\par
\endgroup

\section{Atomic Parameters}
\label{app:atomic}

\begin{table}[H]
\caption{Ionisation potentials of relevant ionic species.}
\label{tab:IPs}
\centering
\begin{tabular}{lcc}
\hline\hline
Ion & IP inf [eV] & IP sup [eV] \\
\hline
O$^{+}$   & 13.6 & 35.1 \\
S$^{+}$   & 10.4 & 23.3 \\
N$^{+}$   & 14.5 & 29.6 \\
O$^{2+}$  & 35.1 & 54.9 \\
Ne$^{2+}$ & 40.9 & 63.4 \\
Ar$^{2+}$ & 27.6 & 40.7 \\
C$^{2+}$  & 24.4 & 47.9 \\
S$^{2+}$  & 23.3 & 34.8 \\
\hline
\end{tabular}
\tablefoot{Ionisation potentials are taken from \citet{osterbrock_astrophysics_2006}. For each ion, we report both the lower IP, corresponding to the energy required to produce that ionisation stage from the preceding one, and the upper IP, corresponding to the energy required to further ionise it to the next stage.}
\end{table}

\begingroup
\centering
\captionof{table}{Atomic data and collision strengths adopted in this work.}
\label{tab:atomic_data}

\begin{tabular}{lcc}
\hline\hline
Ion & Atomic data & Collision strengths \\
\hline
O$^{2+}$  & \citet{Tayal_2017} & \citet{palay_improved_2012} \\
S$^{2+}$  & \citet{Froese_2006} cited by \citet{Podobedova_2009} & \citet{1999ApJ...526..544T} \\
S$^{+}$    & \citet{Rynkun_2019}  & \citet{2010ApJS..188...32T} \\
O$^{+}$    & \citet{10.1093/mnras/198.1.111}; \citet{Wiese_1996} & \citet{Kisielius_2009} \\
Ne$^{2+}$  & \citet{Galavis_1997} &  \citet{Brendan_2000} \\
Ar$^{2+}$  & \citet{Munoz_Burgos_2009} & \citet{Munoz_Burgos_2009}\\
C$^{2+}$  & \citet{Glass_1983, Nussbaumer_1978, Wiese_1996} & \citet{Berrington_1985} \\
N$^{+}$    & \citet{Froese_2004} & \citet{Tayal_2011} \\
\hline
\end{tabular}
\par
\endgroup

\twocolumn

\section{Alternative sulfur diagnostics}
\label{app:Sulfur}

Here we explore alternative strong-line calibrations for S/O. As discussed in Section~\ref{sec:calibrations}, sulfur abundance determinations can be affected by dust depletion and uncertainties in the ionisation correction. In particular, while the available spectra provide access to S$^{+}$ and S$^{2+}$ through the \SII\ and \SIII\ lines, respectively, they do not directly constrain S$^{3+}$, whose [S~IV] emission line at 10.5~$\mu$m lies outside their wavelength coverage. This ion may contribute to the total sulfur abundance in highly ionised nebulae \citep{izotov_chemical_2006, amayo_ICFs_2020}, since the energy required to ionise S$^{2+}$ to S$^{3+}$ is approximately 34.9~eV, nearly identical to the 35.1~eV required to ionise O$^{+}$ to O$^{2+}$ \citep{berg_4zones_2021}. It is therefore useful to assess how different combinations of sulfur and oxygen lines trace S/O.

\begin{figure}
    \centering
    \includegraphics[width=\linewidth]{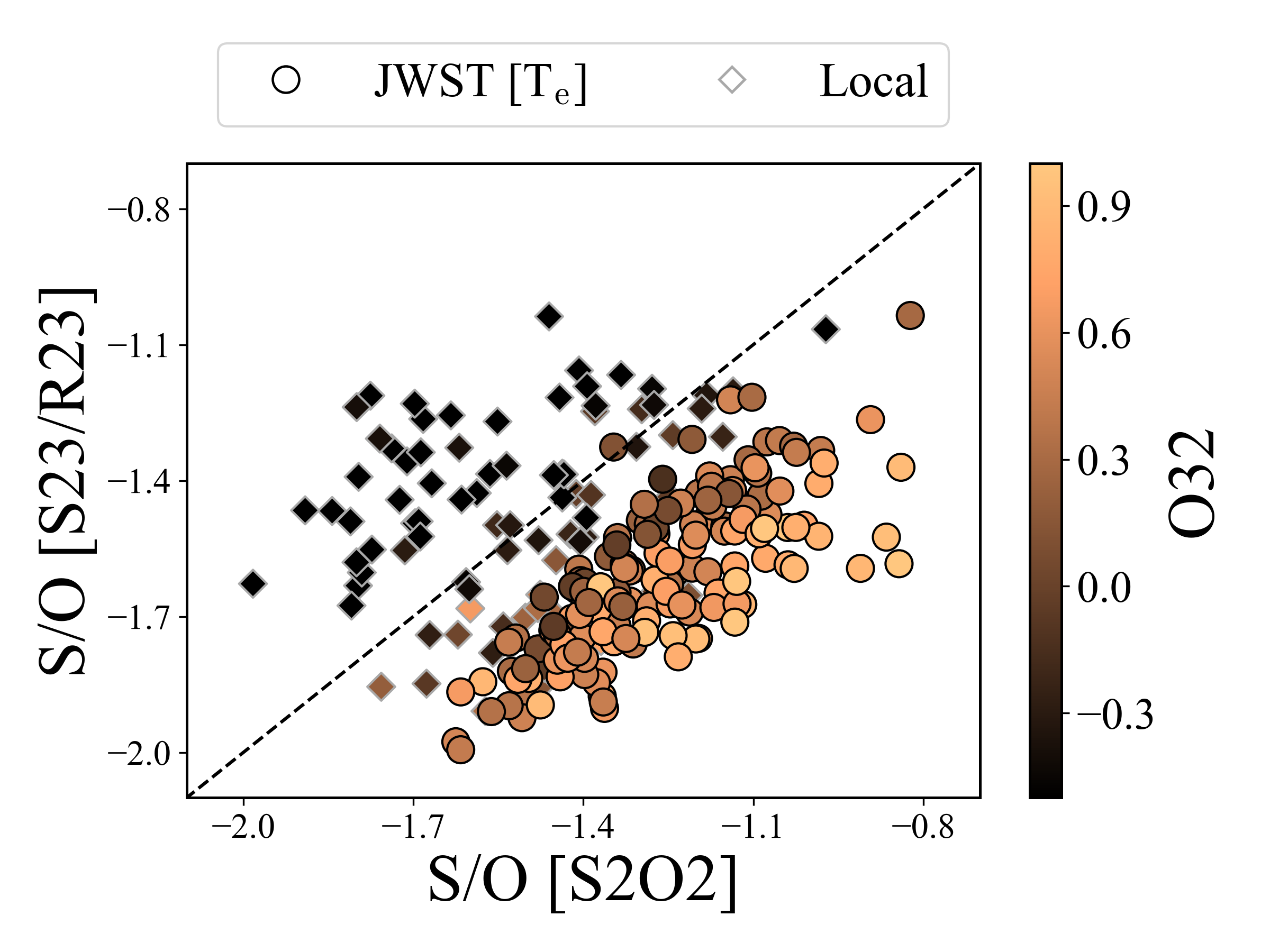}
    \caption{Comparison of $\log(\mathrm{S/O})$ derived from S23/R23 (y-axis) and S2O2 (x-axis). Hexagons, triangles, and circles represent the MARTA, JWST, and local comparison samples, respectively. Symbols are colour-coded by O32, as a proxy for the ionisation parameter.}
    \label{fig:SO_comparison}
\end{figure}

We first compare the S/O abundances inferred using S2O2, which involves only low-ionisation lines, with those obtained using S23/R23, which includes both observed sulfur and oxygen ionisation stages (Figure~\ref{fig:SO_comparison}). The two estimates show a systematic offset: in the high-redshift sample, S23/R23 yields lower S/O values than S2O2, whereas the local comparison sample exhibits the opposite behaviour. For both samples, the departures from the one-to-one relation appear to depend on excitation, as traced by O32. In particular, the highly excited systems, predominantly belonging to the high-redshift sample, show substantially lower S/O when inferred using S23/R23.

These differences may reflect limitations of the adopted ionisation corrections. Over the range of
$W = \mathrm{O}^{+}/(\mathrm{O}^{+}+\mathrm{O}^{2+})$
probed by our samples, both corrections based on \citet{amayo_ICFs_2020} vary weakly. However, the correction applied to S2O2 has a higher normalisation, accounting for the contribution of ionisation stages not traced by the low-ionisation lines. By contrast, the correction applied to S23/R23 is smaller, since this diagnostic already includes S$^{+}$ and S$^{2+}$. This difference in normalisation contributes to the offset between the inferred abundances, but does not by itself explain its dependence on excitation or its reversal between the local and high-z samples.
A possible explanation is that S$^{3+}$ contributes more substantially in the highly excited high-redshift systems than is accounted for by the adopted correction to S23/R23. A harder ionising spectrum could increase the fraction of sulfur in this unobserved stage, leading to an underestimate of S/O from the observed S$^{+}$ and S$^{2+}$ abundances if the correction is insufficient. The excitation-dependent discrepancy is consistent with this interpretation, although the comparison alone cannot establish the S$^{3+}$ contribution or determine which diagnostic provides the more accurate total abundance.

\begin{figure}
    \centering
    \includegraphics[width=\linewidth]{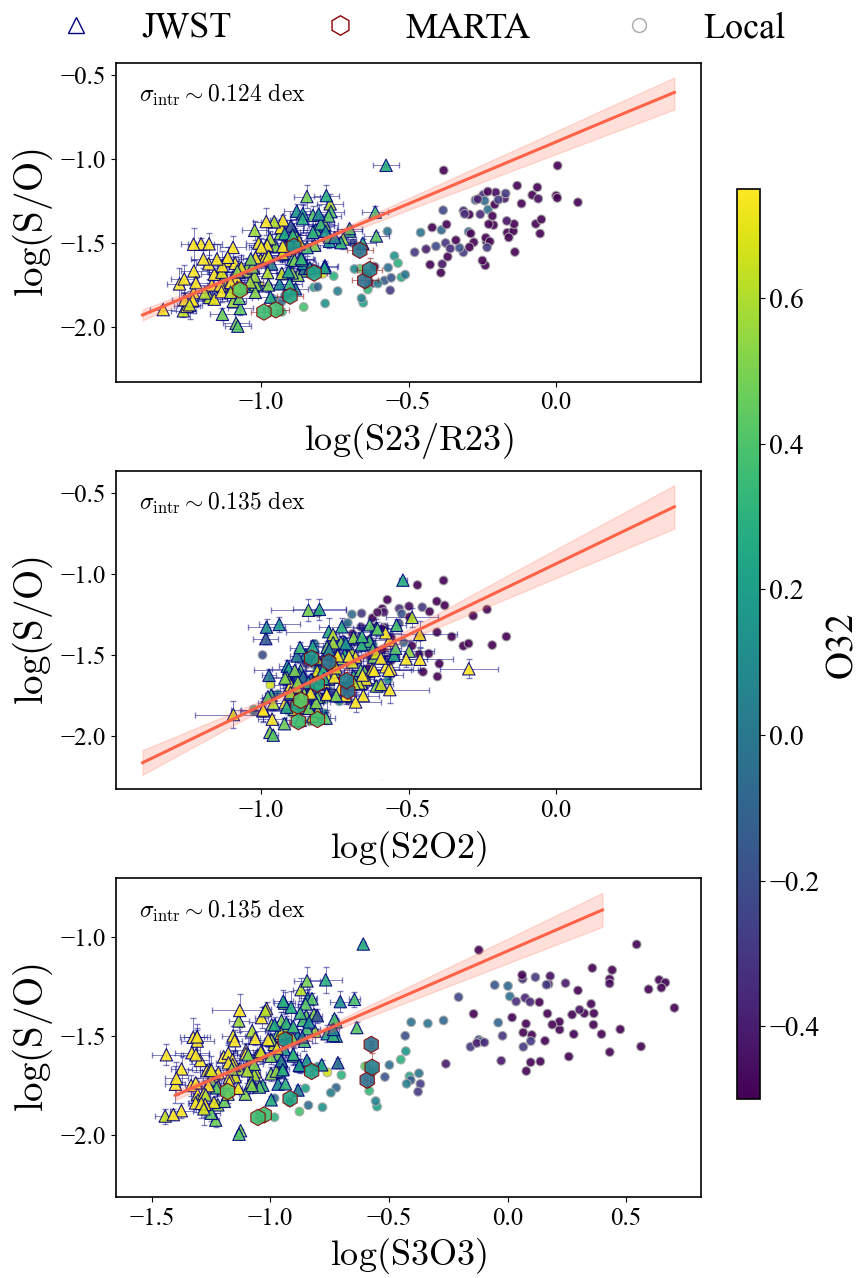}
    \caption{S/O as a function of S23/R23 (top panel), S2O2 (middle panel), and S3O3 (bottom panel). MARTA galaxies are shown as hexagons, the JWST sample as triangles, and local comparison galaxies as small circles. Points are colour-coded by log(O32). The best-fitting linear relations for the combined MARTA and JWST samples, obtained using LinMix, are shown in red; the shaded regions indicate the 16th--84th percentile intervals of the inferred mean relations.}
    \label{fig:S_diagnostics}
\end{figure}

We next compare three strong-line diagnostics, S23/R23, S2O2, and S3O3, which involve either both observed sulfur ionisation stages or each stage individually. Their relations with S/O are shown in Figure~\ref{fig:S_diagnostics}. In all three cases, the reference S/O abundances used to calibrate the diagnostics are derived from both the \SII\ and \SIII\ lines, together with the adopted ionisation correction. For this comparison, we adopt linear relations without the additional dependence on O32 included in the fiducial calibration. The points are nevertheless colour-coded by $\log(\mathrm{O32})$ to illustrate how excitation contributes to the observed dispersion.

We fit the combined MARTA and JWST samples using \texttt{LinMix}, adopting the functional form
\begin{equation}
    \log(\mathrm{S/O}) =
    \alpha + \beta\,\log(\mathrm{ratio}),
\end{equation}
and accounting for measurement uncertainties in both coordinates and intrinsic scatter in $\log(\mathrm{S/O})$. The local comparison sample is shown for reference but is not included in the fits. The resulting coefficients and intrinsic scatters are listed in Table~\ref{tab:alternative_sulfur_calibrations}.

These relations provide alternative estimates of S/O. In particular, S2O2 and S3O3 may be useful when only \SII\ or \SIII, respectively, is available, together with the corresponding oxygen lines. Their intrinsic scatters of approximately $0.12$--$0.14$~dex should be taken into account when applying these calibrations.
The colour coding also reveals a pronounced dependence on O32 in the S3O3 and S23/R23 relations, suggesting that including an explicit excitation term can improve these calibrations, as adopted for our fiducial relation. By contrast, S2O2 shows no similarly clear dependence on O32, although appreciable scatter remains.

Finally, it's worth noticing that the MARTA galaxies occupy an intermediate locus between the local comparison sample and the highest-z systems, both in excitation and in the relative S/O estimates obtained from different diagnostics. This suggests that the discrepancy between the diagnostics may vary continuously with ionisation conditions, with typical Cosmic Noon galaxies bridging the two regimes. Such a pattern would be consistent with a gradual change in the ionic fractions and, potentially, in the hardness of the ionising spectrum, rather than a sharp distinction between local and high-z systems.

\begin{table}
    \centering
    \caption{Coefficients of the alternative sulfur calibrations,
    expressed as $\log(\mathrm{S/O}) =
    \alpha + \beta\,\log(\mathrm{ratio})$.
    The intrinsic scatter, $\sigma_{\rm int}$, is given in dex.}
    \label{tab:alternative_sulfur_calibrations}
    \begin{tabular}{lccc}
        \hline\hline
        Ratio & $\alpha$ & $\beta$ & $\sigma_{\rm int}$ \\
        \hline
        S23/R23 & $-0.94 \pm 0.09$ & $0.88 \pm 0.02$ & 0.124 \\
        S2O2    & $-0.89 \pm 0.07$ & $0.74 \pm 0.07$ & 0.135 \\
        S3O3    & $-1.07 \pm 0.06$ & $0.52 \pm 0.06$ & 0.135 \\
        \hline
    \end{tabular}
\end{table}

\section{Modelling observational selection effects}
\label{app:model}

The abundance trends presented in Section~\ref{sec:abundances} are based on a very heterogeneous sample, from a variety of observing programs (see Appendix \ref{app:jwst_programs}). For a galaxy to appear in any of the four diagnostic planes, its emission lines must be detected above the survey sensitivity threshold. 
Since the S/N cuts on the emission lines effectively impose a selection on the strength of the nebular emission, and hence on the SFR, the resulting sample may be biased with respect to the underlying galaxy population.

Here, we describe how we developed a forward-modelling framework to quantify the impact of selection effects on each of the four abundance diagnostics, and to assess which elements provide the most robust constraints on intrinsic chemical enrichment.

\subsection{Forward model description}

We construct a synthetic population of galaxies at a fixed redshift over a grid of stellar masses spanning $10^7$-$10^{11}$~M$_\odot$ sampled uniformly in logarithmic space with bin width of 0.1 dex. This range is slightly broader than that of our observational sample, where the bulk of our objects reside at M=7.5--10.5 M$_{\odot}$. We draw 300 Monte Carlo realisations per mass bin. For each physical quantity derived step-by-step in the code, we explicitly account for uncertainties by perturbing the value predicted by the various scaling relation discussed below with Gaussian noise, centred on the relation and scaled by its intrinsic scatter.

\paragraph{Star-formation rate.}
Each galaxy is assigned a star formation rate according to the empirical MS of \citet{DiCesare2026}, calibrated on star forming galaxies at $z\sim4$--5, which gives a better representation to our sample than the main sequence presented by \citet{popesso_SFMS_2023}.
An intrinsic scatter of $\sigma_{\rm MS}=0.37$ dex, as reported by \citet{DiCesare2026}, is included.

\paragraph{Gas-phase metallicity.}
Oxygen abundances are drawn from the mass--metallicity relation (MZR). 
The parametrisation is taken from \citet{curti_jades_mzr_2024}, adopting the $z=3$--6 relation over the $z=6$--10 one, as it better matches the redshift distribution of our sample. The relation has a reported intrinsic scatter of 0.07 dex.
As a consistency check, we also implemented the MZR from \citet{sanders_mosdef_mzr_2021}, adopting the $z=3.3$ relation over the $z=2.3$ one for the same reason. We find consistent results between the two MZR prescriptions.

\paragraph{Emission-line fluxes.}
The intrinsic \Hbeta\ flux is derived from the SFR via the \citet{kennicutt_star_1998} calibration (renormalised to a Chabrier IMF).
The \OII\ and \OIIIopt emission-line fluxes are predicted from the gas-phase metallicity O/H using the high-redshift strong-line calibrations of \citetalias{Cataldi2025} for R2 and R3, respectively. We also tested the calibrations of \citet{Sanders_2026_calib} and found that adopting them does not produce any significant differences in the resulting trends.

\begin{figure}
    \centering
    \includegraphics[width=\linewidth]{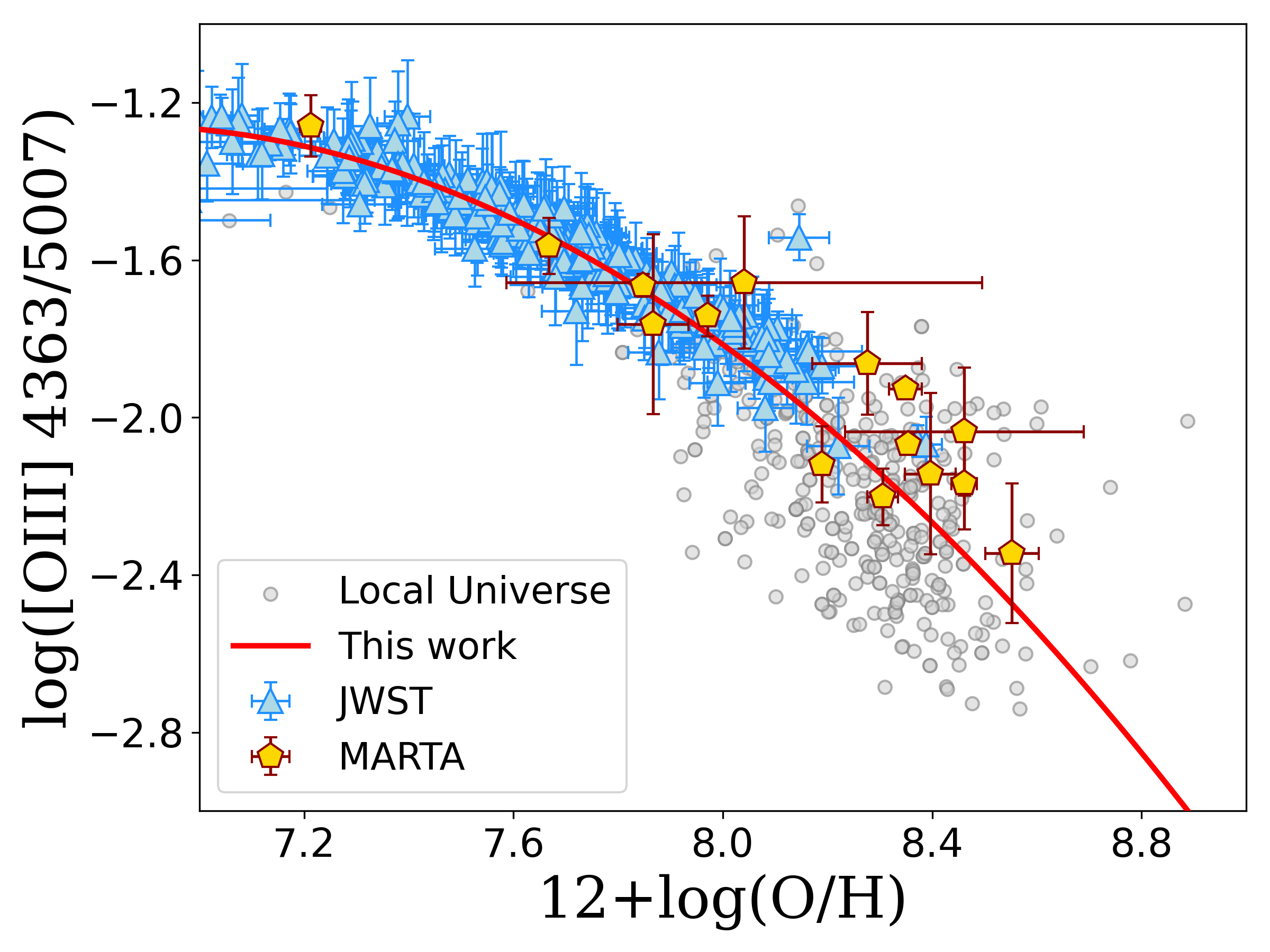}
    \caption{Relation between the \OIIIopt\ auroral-to-nebular line ratio, \OIIIopt$\lambda$4363/\OIIIopt$\lambda$5007, and the gas-phase metallicity. Gold hexagons denote the MARTA galaxies, blue triangles represent the JWST sample described in Section~\ref{sec:DAWN}, and the local reference sample (Section \ref{sec:local}) is shown in grey. The red curve shows the best-fitting quadratic relation.}
    \label{fig:auroral_calib}   
\end{figure}

Finally, the \OIIIopt$\lambda$4363 auroral-line flux is predicted from the galaxy metallicity by adopting a recalibrated relation between the \OIIIopt$\lambda$4363/\OIIIopt$\lambda$5007 line ratio and O/H. This relation has been derived in this work using the compilation of direct-method measurements described in Section~\ref{sec:data}, including both low- and high-$z$ datasets. The resulting calibration is shown in Figure~\ref{fig:auroral_calib}, fitted with a second-order polynomial of the form:
\begin{equation}
    \log\left(\frac{\OIIIopt\lambda4363}{\OIIIopt\lambda5007}\right)
    = a\,x^2 + b\,x + c,
\end{equation}
where $x = 12+\log(\mathrm{O/H})-8.69$. 
The best-fitting coefficients are
\begin{align*}
a &= -0.56 \pm 0.05,\\
b &= -1.79 \pm 0.08,\\
c &= -2.79 \pm 0.03.
\end{align*}

The abundance ratios X/O for each element are assigned from local empirical $\log(\mathrm{X/O})$--metallicity relations (see below), and the corresponding emission-line fluxes are obtained by inverting the strong-line calibrations of Section~\ref{sec:calibrations}.

\paragraph{Abundance ratio and metallicity relations.}

We use the local abundance relations as a baseline for the intrinsic properties of the synthetic galaxies. Specifically, we assume that, before observational selection, galaxies at high redshift follow the same $\log(\mathrm{X/O})$--$12+\log(\mathrm{O/H})$ relations as the local calibration sample described in Section~\ref{sec:local}. We then apply the observational selection to this synthetic population and ask whether it can reproduce the trends measured in our high-z sample. This test therefore assesses the effect of selection under the assumption of no intrinsic evolution in the abundance relations.

For each element, we derive the baseline relation from robust linear fits to the local sample in metallicity bins of width 0.3~dex. For nitrogen, we adopt the relation from \citetalias{Cataldi2026}:
\begin{equation}
\log(\mathrm{N/O}) = Z_0 + \frac{\gamma}{\beta}
  \log\!\left[1 + \left(\frac{\mathrm{O/H}}{(\mathrm{O/H})_0}\right)^\beta\right],
\end{equation}
with $Z_0=-1.385$, $(\mathrm{O/H})_0=4\times10^{-4}$,
$\gamma=2.8$, $\beta=2.3$, and an intrinsic scatter of 0.2~dex.

For sulphur, argon, and neon, we adopt the recent calibrations of \citet{Scholte2026_elements}, which are derived from a large, statistical sample of $\sim$ 49,000 local galaxies and therefore provide a suitable reference for comparison with our high-z measurements.

\paragraph{Dust attenuation.}
Observed fluxes are obtained from intrinsic values by applying a \citet{cardelli_relationship_1989} extinction law ($R_V = 3.1$) with $E(B-V)$ drawn from the mass--$E(B-V)$ relation of \citet{Woodrum2025}. The scatter on $E(B-V)$ is propagated from both the intrinsic dispersion of the relation ($\sigma_{E(B-V)} = 0.13$) and the uncertainty in stellar mass ($\sigma_{\log M_\star} = 0.37$). Each line is attenuated at its rest-frame wavelength independently.

\subsection{Detection thresholds}
\label{sec:thresholds}

Detection thresholds for JWST/NIRSpec are computed following the approach of \citet{Zhu2025}, adapted to the medium-resolution gratings. We use the publicly available NIRSpec sensitivity curves provided by the JWST User Documentation\footnote{\url{https://jwst-docs.stsci.edu/jwst-near-infrared-spectrograph/nirspec-performance/nirspec-sensitivity}}, which give the continuum flux density required to achieve a signal-to-noise ratio of 10 in a 10 ks exposure \citep[see also][]{Giardino2022_nirspec}. 
We assume that the emission lines are spectrally unresolved, such that their observed width is set by the instrumental resolution\footnote{This is appropriate for the majority of star-forming galaxies observed with the NIRSpec medium-resolution gratings, whose intrinsic nebular line widths are typically smaller than the instrumental profile.}.
The corresponding $1\sigma$ noise level for a 10 ks exposure is therefore
$\sigma(10\,{\rm ks})=f_{\lambda}/10$,
which is then rescaled to an arbitrary exposure time $t$ assuming background-limited observations,

\begin{equation}
\sigma(t)=\sigma(10\,{\rm ks})
\sqrt{\frac{10\,{\rm ks}}{t}}.
\end{equation}

The minimum detectable integrated line flux at a significance level of $N\sigma$ is then $F_{\rm lim}=N\,\sigma(t)\,\Delta\lambda$, where we adopt $N=3$. Following the assumption of spectrally unresolved line, the observed line width $\Delta\lambda$ is $
\Delta\lambda \simeq \lambda_{\rm obs}/R(\lambda_{\rm obs}) $,
where $R(\lambda_{\rm obs})$ is the wavelength-dependent spectral resolving power of NIRSpec. For each emission line, we evaluate this width at the wavelength that gives the most conservative flux limit within the relevant grating coverage, using the tabulated resolution curves for G140M and G235M from the JWST User Documentation\footnote{\url{https://jwst-docs.stsci.edu/jwst-near-infrared-spectrograph/nirspec-instrumentation/nirspec-dispersers-and-filters}}. 

\begin{figure}
    \centering
    \includegraphics[width=\linewidth]{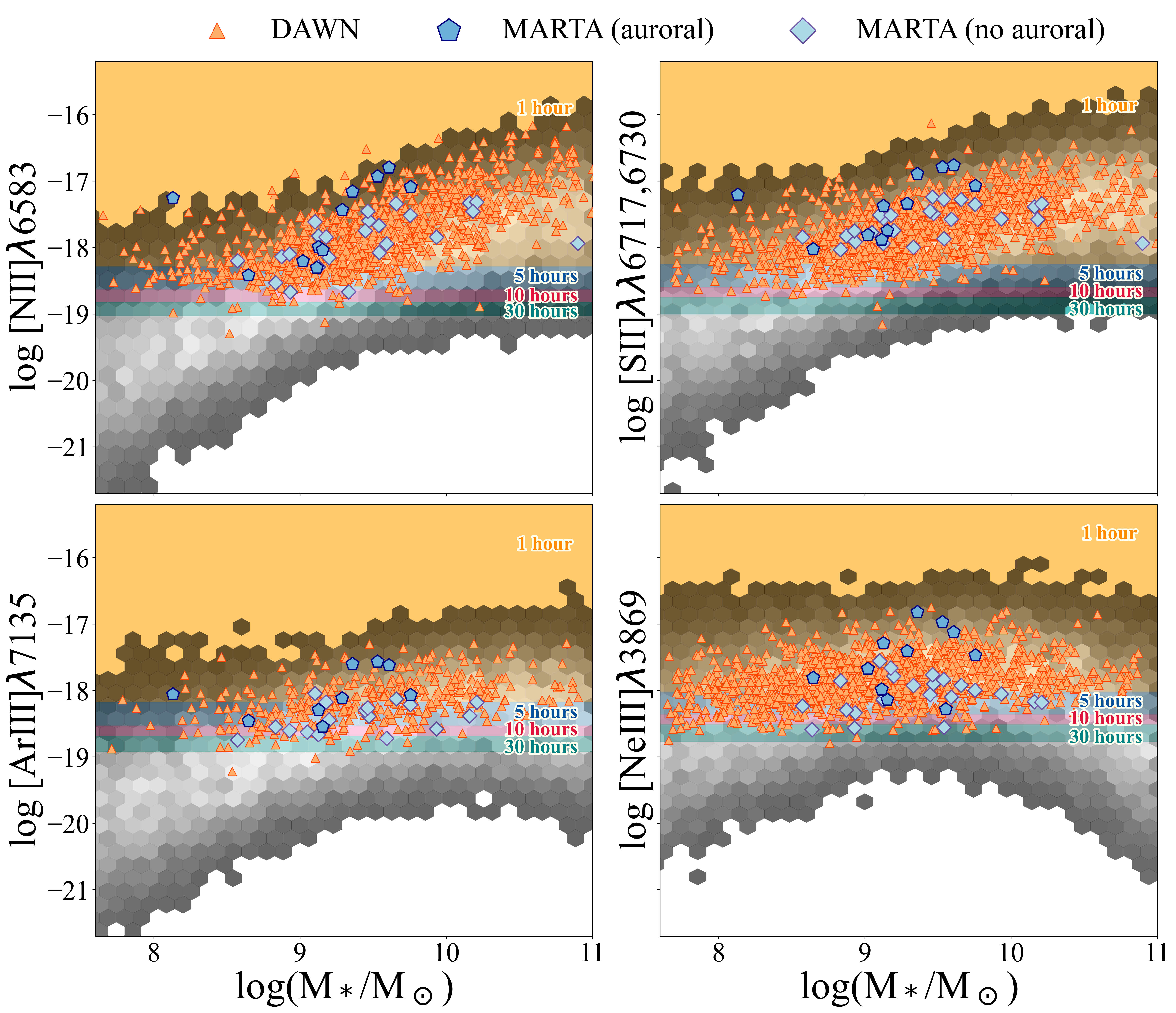}
    \caption{Simulated flux distributions (hexbin density maps) of
    \NII$\lambda$6583, \SII$\lambda\lambda$6716,6730, \ArIII$\lambda$7135, and
    \NeIII$\lambda$3869 as a function of stellar mass. Observed fluxes from our high-redshift sample are shown as individual points. JWST/NIRSpec $3\sigma$ detection thresholds at the observed wavelength of each line are shown for integration times of 1, 5, 10, and 30 hours.}
    \label{fig:fluxes_masses}
\end{figure}

With this information, we  compute the expected signal-to-noise values for \NII$\lambda$6583, \SII$\lambda\lambda$6716,6730, \ArIII$\lambda$7135, \NeIII$\lambda$3869 and \OIIIopt$\lambda$4363 emission lines, assuming representative integration times of 1, 5, 10, and 30 hours, as shown in Figure \ref{fig:fluxes_masses}. The resulting flux limits are also reported in Table \ref{tab:fluxes}.
We do not perform this analysis for carbon, as the currently observed sample is too small to draw robust conclusions about any trends that might emerge in the simulated population.

Figure~\ref{fig:fluxes_masses} provides a sanity check of our forward model: the simulated flux distributions broadly reproduce the observed line fluxes, and the adopted detectability thresholds are consistent with the region occupied by the observed sample. This agreement supports the use of the simulated population to assess how line detectability affects the observed abundance distributions.

\begin{table}[htbp]
\centering
\caption{Flux detection thresholds (S/N > 3) for the emission lines considered in this work.}
\label{tab:fluxes}
\setlength{\tabcolsep}{5pt}
\begin{tabular}{lcccc}
\hline\hline
Line & 1h & 5h & 10h & 30h \\
\hline
\NII$\lambda$6583  & 5.6 & 2.5 & 1.8 & 1.0 \\
\SII$\lambda\lambda$6716,6730 &  5.1 & 2.3 & 1.6 & 0.9 \\
\ArIII$\lambda$7135  & 6.6 & 2.9 & 2.1 & 1.2 \\
\NeIII$\lambda$3869  & 9.6 & 4.3 & 3.0 & 1.8 \\
\OIIIopt$\lambda$4363 & 8.1 & 3.7 & 2.5 & 1.6 \\
\hline
\end{tabular}
\tablefoot{All fluxes are in units of $10^{-19} \, \mathrm{erg/s/cm}^2$.}
\end{table}

\end{appendix}

\end{document}

%% file: config/macros.tex
\usepackage{etoolbox}
\makeatletter
\newcommand\sendemail[3]{
\edef\@tempa{mailto:#1?subject=#2 }%
\edef\@tempb{\expandafter\html@spaces\@tempa\@empty}%
\href{\@tempb}{#3}}

\catcode\%=11
\def\html@spaces#1 #2{#1
\catcode\%=14
\makeatother

\newcommand{\citationneeded}{\textcolor{ForestGreen}{$^{\rm citation\;needed}$}}
\let\oldtextsigma\textsigma
\renewcommand{\textsigma}{\oldtextsigma\xspace}
\let\oldtextalpha\textalpha
\renewcommand{\textalpha}{\oldtextalpha\xspace}
\let\oldAA\AA
\renewcommand{\AA}{\text{\oldAA}\xspace}
\let\oldtextdegree\textdegree
\renewcommand{\textdegree}{\oldtextdegree\xspace}

\newcommand{\orcid}[2]{\href{http://orcid.org/#2}{#1{\includegraphics[height=10pt]{config/orcid.pdf}}}}

\newcommand{\kms}{\ensuremath{\mathrm{km\,s^{-1}}}\xspace}
\newcommand{\MSun}{\ensuremath{{\rm M}_\odot}\xspace}
\newcommand{\yr}{\ensuremath{{\rm yr}}\xspace}
\newcommand{\Myr}{\ensuremath{{\rm Myr}}\xspace}
\newcommand{\Gyr}{\ensuremath{{\rm Gyr}}\xspace}
\newcommand{\peryr}{\ensuremath{{\rm yr^{-1}}}\xspace}
\newcommand{\Lsun}{\hbox{\,${\rm L}_\odot$}}
\newcommand{\mum}{\text{\textmu m}\xspace}
\newcommand{\dex}{\text{dex}\xspace}
\newcommand{\kpc}{\text{kpc}\xspace}
\newcommand{\ZH}{\text{[Z/H]}\xspace}
\newcommand{\CO}{\text{[C/O]}\xspace}
\newcommand{\NO}{\text{[N/O]}\xspace}
\newcommand{\FeH}{\text{[Fe/H]}\xspace}
\newcommand{\percm}[1]{\ensuremath{\rm cm^{#1}}\xspace}

\newcommand{\eps}{\ensuremath{\epsilon}\xspace}
\newcommand{\mstar}{\ensuremath{M_\star}\xspace}
\newcommand{\mgas}{\ensuremath{M_\mathrm{gas}}\xspace}
\newcommand{\re}{\ensuremath{R_\mathrm{e}}\xspace}
\newcommand{\NH}{\ensuremath{N_\mathrm{H}}\xspace}
\newcommand{\tauv}{\ensuremath{\tau_\mathrm{V}}\xspace}
\newcommand{\AV}{\ensuremath{A_\mathrm{V}}\xspace}
\newcommand{\xid}{\ensuremath{\xi_\mathrm{d}}\xspace}
\newcommand{\logoh}{\ensuremath{12 + \log\,(\mathrm{O/H})}\xspace}

\newcommand{\nelec}{\ensuremath{n_\mathrm{e}}\xspace}
\newcommandx{\Mout}[2][1=,2=]{\ensuremath{M_{\mathrm{out}{#2}}^{#1}}\xspace}
\newcommandx{\Mdotout}[2][1=,2=]{\ensuremath{\dot{M}_{\mathrm{out}{#2}}^{#1}}\xspace}

\newcommandx{\fluxdcgs}[1][1=-20]{\ensuremath{\mathrm{10^{#1}~erg~s^{-1}~cm^{-2}~\AA^{-1}}}\xspace}
\newcommandx{\fluxcgs}[1][1=-20]{\ensuremath{\mathrm{10^{#1}~erg~s^{-1}~cm^{-2}}}\xspace}
\newcommandx{\powercgs}[1][1=44]{$\times 10^{#1}$~erg~s$^{-1}$\xspace}
\newcommand{\Av}{\ensuremath{A_V}\xspace}

\newcommand{\Te}{\ensuremath{T_\text{e}}\xspace}
\newcommand{\Tiii}{\Te[\ion{O}{iii}]\xspace}
\newcommand{\Tii}{\Te[\ion{O}{ii}]\xspace}
\newcommand{\TSii}{\Te[\ion{S}{ii}]\xspace}
\newcommand{\TSiii}{\Te[\ion{S}{iii}]\xspace}

\newcommand{\Ne}{n$_{\text{e}}$\xspace}

\newcommand{\EWr}{EW$_{\text{0}}$\xspace}

\newcommand{\jwst}{\textit{JWST}\xspace}
\newcommand{\hst}{\textit{HST}\xspace}
\newcommand{\ppxf}{{\sc ppxf}\xspace}
\newcommand{\beagle}{{\sc beagle}\xspace}
\newcommand{\forcepho}{{\sc forcepho}\xspace}
\newcommand{\prospector}{{\sc prospector}\xspace}

%% file: config/emlines.tex
\newcommand{\Lyalpha}{\text{Ly\textalpha}\xspace}
\newcommand{\Halpha}{\text{H\textalpha}\xspace}
\newcommand{\Hbeta}{\text{H\textbeta}\xspace}
\newcommand{\Hgamma}{\text{H\textgamma}\xspace}
\newcommand{\Hdelta}{\text{H\textdelta}\xspace}
\newcommand{\Pabeta}{\text{Pa\textbeta}\xspace}
\newcommand{\Hepsilon}{\text{H\textepsilon}\xspace}

\newcommandx{\permittedEL}[6][1=O,2=III,3=,4=,5=,6=]{\text{{#1}\,{\sc {#2}}{#3}{#4}{#5}{#6}}\xspace}
\newcommandx{\semiforbiddenEL}[6][1=O,2=III,3=,4=,5=,6=]{\text{{#1}\,{\sc{#2}}]{#3}{#4}{#5}{#6}}\xspace}
\newcommandx{\forbiddenEL}[6][1=O,2=III,3=,4=,5=,6=]{\text{[{#1}\,{\sc{#2}}]{#3}{#4}{#5}{#6}}\xspace}

\newcommand{\EW}[1]{\text{EW(#1)}\xspace}

\newcommand{\hii}{\permittedEL[H][ii]}

\newcommand{\NV}{\permittedEL[N][v]}
\newcommandx{\NVL}[1][1=1243]{\permittedEL[N][v][\textlambda][#1]}
\newcommandx{\NVall}{\permittedEL[N][v][\textlambda][\textlambda][1239,][1243]}
\newcommand{\NII}{\forbiddenEL[N][ii]}

\newcommand{\NIV}{\semiforbiddenEL[N][iv]}
\newcommandx{\NIVL}[1][1=1486]{\semiforbiddenEL[N][iv][\textlambda][#1]}

\newcommand{\CIV}{\permittedEL[C][iv]}
\newcommandx{\CIVL}[1][1=1550]{\permittedEL[C][iv][\textlambda][#1]}
\newcommand{\CIVall}{\permittedEL[C][iv][\textlambda][\textlambda][1549,][1551]}

\newcommand{\HeII}{\permittedEL[He][ii]}
\newcommandx{\HeIIL}[1][1=1640]{\permittedEL[He][ii][\textlambda][#1]}

\newcommand{\OIII}{\semiforbiddenEL[O][iii]}
\newcommandx{\OIIIL}[1][1=1666]{\semiforbiddenEL[O][iii][\textlambda][#1]}
\newcommand{\OIIIall}{\semiforbiddenEL[O][iii][\textlambda][\textlambda][1661,][1666]}
\newcommand{\SiIII}{\semiforbiddenEL[Si][iii]}

\newcommand{\OIIIopt}{\forbiddenEL[O][iii]}
\newcommandx{\OIIIoptL}[1][1=5007]{\forbiddenEL[O][iii][\textlambda][#1]}
\newcommand{\OIIIoptall}{\forbiddenEL[O][iii][\textlambda][\textlambda][4959,][5007]}

\newcommand{\NIII}{\semiforbiddenEL[N][iii]}
\newcommandx{\NIIIL}[1][1=1750]{\semiforbiddenEL[N][iii][\textlambda][#1]}
\newcommand{\NIIIall}{\semiforbiddenEL[N][iii][\textlambda][\textlambda][1747--][1754]}

\newcommandx{\CIII}{\semiforbiddenEL[C][iii]}
\newcommandx{\CIIIL}[1][1=1909]{\semiforbiddenEL[C][iii][\textlambda][#1]}
\newcommand{\CIIIall}{\semiforbiddenEL[C][iii][\textlambda][\textlambda][1907,][1909]}

\newcommand{\CIIIp}{\permittedEL[C][iii]}
\newcommand{\NIIIp}{\permittedEL[N][iii]}

\newcommand{\NeIV}{\forbiddenEL[Ne][iv]}
\newcommandx{\NeIVL}[1][1=2424]{\forbiddenEL[Ne][iv][\textlambda][#1]}
\newcommand{\NeIVall}{\forbiddenEL[Ne][iv][\textlambda][\textlambda][2422,][2424]}

\newcommand{\MgII}{\permittedEL[Mg][ii]}
\newcommandx{\MgIIL}[1][1=2803]{\permittedEL[Mg][ii][\textlambda][#1]}
\newcommand{\MgIIall}{\permittedEL[Mg][ii][\textlambda][\textlambda][2796,][2803]}

\newcommand{\NeV}{\forbiddenEL[Ne][v]}
\newcommandx{\NeVL}[1][1=3426]{\forbiddenEL[Ne][v][\textlambda][#1]}
\newcommand{\NeVall}{\forbiddenEL[Ne][v][\textlambda][\textlambda][3346,][3426]}

\newcommand{\OII}{\forbiddenEL[O][ii]}
\newcommandx{\OIIL}[1][1=3727]{\forbiddenEL[O][ii][\textlambda][#1]}
\newcommand{\OIIall}{\forbiddenEL[O][ii][\textlambda][\textlambda][3726,][3729]}

\newcommand{\OIIaur}{\forbiddenEL[O][ii][\textlambda][\textlambda][7320,][7330]}

\newcommand{\SII}{\forbiddenEL[S][ii]}
\newcommandx{\SIIL}[1][1=6725]{\forbiddenEL[S][ii][\textlambda][#1]}
\newcommand{\SIIall}{\semiforbiddenEL[S][ii][\textlambda][\textlambda][6718,][6732]}

\newcommand{\SIII}{\forbiddenEL[S][iii]}
\newcommandx{\SIIIL}[1][1=9068]{\forbiddenEL[S][iii][\textlambda][#1]}

\newcommand{\SIIIall}{\semiforbiddenEL[S][iii][\textlambda][\textlambda][9068,][9532]}

\newcommand{\NeIII}{\forbiddenEL[Ne][iii]}
\newcommandx{\NeIIIL}[1][1=3869]{\forbiddenEL[Ne][iii][\textlambda][#1]}
\newcommand{\NeIIIall}{\forbiddenEL[Ne][iii][\textlambda][\textlambda][3869,][39xx]}

\newcommand{\FeII}{\forbiddenEL[Fe][ii]
[\textlambda][4359]}

\newcommand{\ArIII}{\forbiddenEL[Ar][iii]}
\newcommand{\ArIIIL}{\forbiddenEL[Ar][iii]
[\textlambda][7135]}

\newcommand{\OI}{\forbiddenEL[O][i]}
\newcommand{\OIL}{\forbiddenEL[O][i]
[\textlambda][6300]}

\newcommand{\hda}{\ensuremath{\mathrm{H\text{\textdelta}_A}}\xspace}
\newcommand{\hga}{\ensuremath{\mathrm{H\text{\textgamma}_A}}\xspace}